\documentclass[prb, aps, twocolumn, groupedaddress, eqsecnum, superscriptaddress, nofootinbib]{revtex4-2}

\usepackage[dvipsnames]{xcolor}
\usepackage{graphicx}
\usepackage[normalem]{ulem} 
\usepackage{enumitem}       

\usepackage{amsmath}
\usepackage{amssymb}
\usepackage{bbm}            
\usepackage{newtxtext,newtxmath} 
\usepackage{bm}             

\usepackage[version=4]{mhchem} 
\usepackage{hhline}
\usepackage{array}

\usepackage[colorlinks=true,linkcolor=Blue,citecolor=Blue,urlcolor=Blue]{hyperref}
\usepackage{cleveref} 

\newcommand{\nn}{\nonumber \\}
\newcommand{\Sp}{{\bf S}}
\newcommand{\XXZ}{{\rm XXZ}}

\newcommand{\kb}{{k_{\rm B}}}

\begin{document}
\title{Spin-orbit-induced quantum chiral phases}

\author{Daesik \surname{Kim}}
\affiliation{Department of Physics, Sungkyunkwan University, Suwon 16419, South Korea}

\author{Hyojae \surname{Jeon}}
\affiliation{Department of Physics, Sungkyunkwan University, Suwon 16419, South Korea}

\author{Seongjun \surname{Park}}
\affiliation{Department of Physics, Korea Advanced Institute of Science and Technology, Daejeon 34141, Korea}

\author{Seungho \surname{Lee}}
\affiliation{Department of Physics, Sungkyunkwan University, Suwon 16419, South Korea}
\affiliation{Institute of Basic Science, Sungkyunkwan University, Suwon 16419, South Korea}

\author{Jung Hoon \surname{Han}}
\email{hanjemme@gmail.com}
\affiliation{Department of Physics, Sungkyunkwan University, Suwon 16419, South Korea}

\author{Hyun-Yong \surname{Lee}}
\email{hyunyong@korea.ac.kr}
\affiliation{Division of Semiconductor Physics, Korea University, Sejong 30019, Korea}
\affiliation{Department of Applied Physics, Graduate School, Korea University, Sejong 30019, Korea}

\begin{abstract} 
The scalar spin chirality (SSC), whose nonzero value $\langle {\bf S}_i \cdot ({\bf S}_j \times {\bf S}_k) \rangle \neq 0$ implies the breaking of time-reversal and certain point-group symmetries in the ground state, is a key quantity characterizing chiral magnetism in both classical and quantum settings. The classical SSC is manifested, for instance, in skyrmion crystal phase, while the quantum SSC is still highly sought after in various frustrated spin-1/2 models. An interesting possibility that has not been explored so far is the case in which SSC is symmetry-wise allowed, yet remains zero classically due to the collinear or coplanar arrangement of spins, but is generated by virtue of quantum fluctuation. We demonstrate the existence of precisely such a phase in a spin-1/2 triangular-lattice model with XXZ interaction, spin-orbit-induced exchange interactions, and an external magnetic field. Using iDMRG, we thoroughly map out the phase diagram of the model and identify several phases with coexisting magnetic order and SSC. The nonzero SSC arises despite the classical magnetic order being collinear or coplanar. We provide detailed magnon analysis to ascribe its origin to quantum fluctuations around classical magnetic order. The estimate of SSC from magnon analysis agrees with iDMRG results quantitatively. We map out the magnon spectrum and its Berry curvature, culminating in the prediction of finite thermal Hall conductivity in these phases with SSC. 
\end{abstract}

\date{\today}
\maketitle

\section{introduction}
Experimental and theoretical search for a new phase of matter in quantum spin systems that goes beyond the framework of conventional magnetic order has been a continuing theme for decades. Well-known instances of such states are the resonating valence bond and chiral spin liquid, characterized by nonzero composite order $\langle \Sp_i \cdot \Sp_j \rangle$ or $\chi_{ijk} = \langle \Sp_i \cdot \Sp_j \times \Sp_k \rangle$ for the adjacent sites $i,j,k$ on the lattice in the absence of magnetic order, \textit{i.e.} $\langle \Sp_j \rangle =0$. Search for concrete evidence for such states is, to say the least, still ongoing. In particular, the non-magnetic state supporting the scalar spin chirality (SSC) $\chi_{ijk}$ breaks both time-reversal and certain point-group symmetries~\cite{WWZ89}. A more modest kind of composite order is the vector spin chirality (VSC) $\langle \Sp_i \times \Sp_j \rangle$, which breaks spatial reflection symmetry but not time-reversal symmetry. In fact, the two spin chiralities are closely related in magnetic systems as $\langle \Sp_i\rangle \neq 0$ together with nonzero VSC could lead to SSC $\chi_{ijk} \sim \langle \Sp_i \rangle \cdot \langle \Sp_j \times \Sp_k \rangle$.
An interesting and readily observable implication of such composite order is the thermal Hall transport in insulating magnets~\cite{katsura10}. 

In this work, we address the question whether the quantum SSC can remain finite even when the corresponding classical spin configuration, which is coplanar, disallows classical SSC. We report numerical and analytical investigation of a two-dimensional $S=1/2$ spin Hamiltonian with XXZ interactions, spin-orbit-coupling (SOC)-induced exchange interactions, and an external magnetic field. The quantum ground state as worked out by iDMRG develops collinear or coplanar magnetic order and, at the same time, a nonzero SSC. In the zero-SOC limit, the SSC is forbidden by an effective time-reversal symmetry of the Hamiltonian, even though the physical time-reversal symmetry is broken by the external field. The combination of SOC and a finite magnetic field lifts all symmetry constraints on the SSC, as detailed in Sec.\,\ref{sec:model}. Consequently, a nonzero SSC can develop without requiring spontaneous symmetry breaking in the ground state.

We have uncovered several magnetically ordered phases in which the classical spin configuration is collinear or coplanar, so that the SSC should vanish on classical consideration alone, yet its full quantum expectation value is nonzero. We attribute this feature to quantum fluctuations and refer to it as quantum SSC (QSSC), to distinguish it from the classical SSC realized, \textit{e.g.}, in skyrmion spin textures. The magnetically ordered state supporting QSSC will be called a {\it quantum chiral magnet}. To elucidate the microscopic origin of QSSC, we perform a Holstein-Primakoff (HP) analysis around each classical spin configuration, gaining quantitative agreement with the SSC values obtained with iDMRG. We further compute the magnon Berry curvature and the thermal Hall conductivity, providing an experimentally accessible signature of the quantum-chiral magnetic phases.

The spin model we study has been under a great deal of scrutiny recently 
due to the appearance of spin-supersolid phases in its phase 
diagram~\cite{Yamamoto14, Li16, Zhu19, zhu2024continuum, xiang2024giant, 
Plat18, liu2024supersolidity, Homeier25}. A spin-supersolid is a quantum 
magnetic phase in which long-range spin-density order and transverse 
superfluid order coexist. Given the growing experimental and theoretical 
interest in these phases, it is a pressing question how they are 
modified or driven into new forms of quantum order when spin-orbit 
coupling is present. In this work, we show that SOC-induced exchange 
interactions indeed give rise to a qualitatively new phenomenon---the 
quantum scalar spin chirality---that is absent in the pure XXZ limit.

\section{Model}
\label{sec:model}
\begin{figure}[]
    \includegraphics[width=0.99\columnwidth]{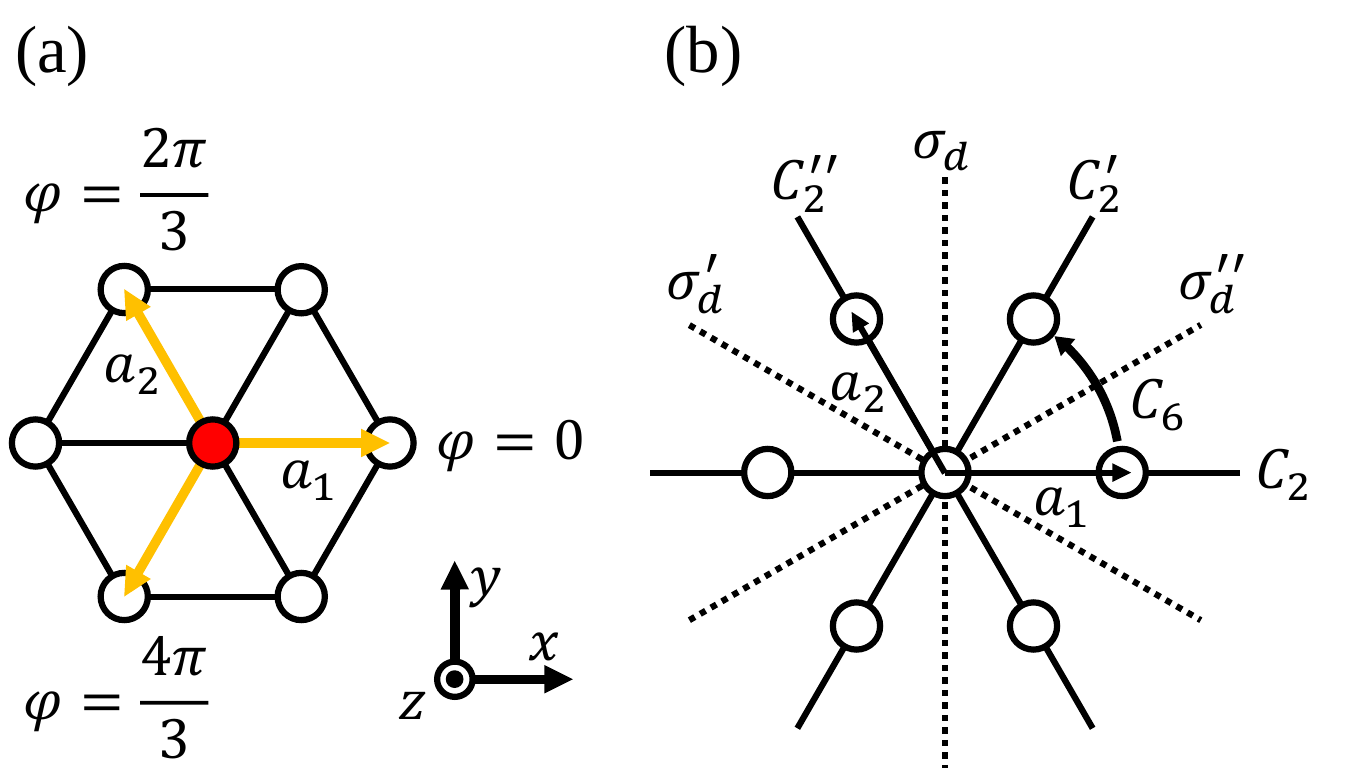}
    \caption{(a) Schematic of the triangular lattice showing the three anisotropic nearest-neighbor bonds in \(H_{\rm PD}\) and \(H_{\Gamma}\)\,[Eq.\,\eqref{eq:hamiltonian}].
    (b) Symmetry operations of the point group $\mathsf{{D}_{3d}}=\{E,i,C_3,C_3^{2},S_6,S_6^{5},C_2,C_2',C_2'',\sigma_d,\sigma_d',\sigma_d''\}$. The solid lines indicate the rotation axes (perpendicular to $z$ axis) of the three $C_2$ rotation operations. The dashed lines indicate the reflection planes (parallel to $z$ axis) of the three $\sigma_{d}$ reflection operations.
    The $S_{6}$ operation indicates the $C_{6}$ rotation followed by a reflection through the $xy$ plane.
    }
    \label{fig:schematic}
\end{figure}

The model under consideration is a quantum spin-1/2 XXZ model on a triangular lattice augmented by anisotropic interactions, given by the Hamiltonian
\begin{align}
    H & = H_{\rm XXZ} \;+\; H_{\rm PD} \;+\; H_{\Gamma}, \nn 
    H_{\rm XXZ}
    &= \sum_{\langle ij\rangle}\!\Big[\, J\big(\hat S_i^{x}\hat S_j^{x}+\hat S_i^{y}\hat S_j^{y}\big)+J_z\,\hat S_i^{z}\hat S_j^{z}\Big]
       \;-\; h_z\sum_i \hat S_i^{z}, \nonumber\\
    H_{\rm PD}
    &= 2J_{\rm PD}\sum_{\langle ij\rangle}
       \Big(\,[\hat S^{x},\hat S^{y}]_{ij}\cos\varphi_{ij}-\{\hat S^{x},\hat S^{y}\}_{ij}\sin\varphi_{ij}\Big), \nonumber\\
    H_{\Gamma}
    &= J_{\Gamma}\sum_{\langle ij\rangle}
       \Big(\,\{\hat S^{y},\hat S^{z}\}_{ij}\cos\varphi_{ij}-\{\hat S^{x},\hat S^{z}\}_{ij}\sin\varphi_{ij}\Big) , 
    \label{eq:hamiltonian}
\end{align}
where $[\hat{S}^\alpha, \hat{S}^\beta ]_{ij} \equiv \hat{S}^{\alpha}_{i} \hat{S}^{\alpha}_{j} - \hat{S}^{\beta}_{i} \hat{S}^{\beta}_{j}$, $\lbrace \hat{S}^\alpha, \hat{S}^\beta \rbrace_{ij} \equiv \hat{S}^{\alpha}_{i} \hat{S}^{\beta}_{j} + \hat{S}^{\beta}_{i} \hat{S}^{\alpha}_{j}$, and the angle \(\varphi_{ij} = 0, \pm \frac{2\pi}{3}\) depends on the orientation of the bond \(\langle ij \rangle\) as illustrated in Fig.\,\ref{fig:schematic} (a). The pseudo-dipolar\,(PD) interaction $H_{\rm PD}$ and $H_\Gamma$ exist generically in strongly spin-orbit-coupled (SOC) QSL candidates on a triangular lattice such as YbMgGaO$_4$~\cite{chen15,chen16}. The Dzyaloshinskii-Moriya interaction is absent as the 3D inversion symmetry $i$ is assumed intact. The XXZ part of the Hamiltonian has the U(1) symmetry under the spin rotation about the $z$-axis.  

Here, we discuss the detailed symmetry properties of the model Hamiltonian, focusing on their impact on the SSC.
In the absence of a magnetic field ($h_z=0$), the Hamiltonian $H$ enjoys full $\mathsf{{D}_{3d}}$ point group symmetry \cite{PhysRevX.9.021017,PhysRevLett.119.157201}.
The external magnetic field explicitly breaks the $\mathsf{{D}_{3d}}$  symmetry, which is reduced to $\mathsf{{S}_{6}}=\{E,i,C_3,C_3^{2},S_6,S_6^{5}\}$; the explicit symmetry notations are illustrated in Fig. \ref{fig:schematic} (b).

It is important to note that three $C_2$ rotations and three $\sigma_d$ reflections change the ordering of $(i,j,k)$, from counterclockwise to clockwise, thereby flipping the sign of $\chi_{ijk}$.
As a result, the $\mathsf{{D}_{3d}}$ symmetry forbids nonzero SSC.
On the other hand, the group $\mathsf{{S}_{6}}$, which is generated by the symmetry operation $S_6$, preserves the sign of $\chi_{ijk}$.
Unlike the group $\mathsf{{D}_{3d}}$, the group $\mathsf{{S}_{6}}$ does not forbid nonzero SSC.

Besides translational and point group symmetry, there is an additional symmetry structure acting only on the spin degree of freedom, summarized in Table. \ref{Tab:Symmetry Structure}.
First, in the zero-SOC limit, the Hamiltonian $H$ has global $U(1)$ rotational symmetry generated by spin rotation along the $z$-axis $U_z(\phi) = \exp(-i \phi \sum_j \hat{S}_j^z )$, anti-unitary $Z_2$ symmetry generated by $\mathcal{T}'= U_y(\pi)\mathcal{T}$, and other anti-unitary $Z_2$ symmetry generated by $\mathcal{T}''= C_{2}\,\mathcal{T}$.
In this case, while the time-reversal symmetry is explicitly broken, an effective time-reversal symmetry ($\mathcal{T}'$) remains and prohibits a nonzero SSC, as noted in similar contexts where an effective time-reversal symmetry forbids a finite SSC~\cite{Esaki25}.
Second, PD interaction breaks the $\mathcal{T}'$ symmetry and reduces $U(1)$ symmetry to $Z_2$ symmetry. 
In this case, the $\mathcal{T}''$ symmetry remains, but it does not prohibit a nonzero SSC.
Third, the ${\rm \Gamma}$ interaction breaks the $\mathcal{T}'$ and $U(1)$ symmetry.
Again, no symmetry constraint forbids a nonzero SSC. 
In the presence of both spin-orbit-induced interactions, as $H_{\rm \Gamma}$ has the lowest symmetry structure, the symmetry structure is dictated by $H_{\rm \Gamma}$.

\begin{table}[h]
    \centering
    \begingroup
    \normalsize
    \renewcommand{\arraystretch}{1.3}
    \begin{tabular}{|c|c|c|}
    \hline
    $H$ & Symmetry Group & SSC\\
    \hline
    $H_{\rm XXZ}$ & $\{U_{z}(\phi)\}\times{}\{E,\mathcal{T}'\}\times{}\{E,\mathcal{T}''\}$ & Forbidden \\
    \hline
    $H_{\rm XXZ}+H_{\rm PD}$ & $\{E,U_{z}(\pi)\}\times{}\{E,\mathcal{T}''\}$ & Allowed \\
    \hline
    $H_{\rm XXZ}+H_{\rm \Gamma}$ & $\{E,\cal{T}''\}$ & Allowed \\
    \hline
    \end{tabular}
    \endgroup
    \caption{Internal symmetry structure of the Hamiltonian for different SOC terms, in the presence of a magnetic field. The translational and point group symmetries are omitted for simplicity.
    }
    \label{Tab:Symmetry Structure}
\end{table}

\section{Phase diagram} 
The classical phase diagram of the model in Eq.\,\eqref{eq:hamiltonian}, in the absence of the Zeeman term, was established in Ref.\,\cite{chen16}, while the quantum phase diagram was further explored in Refs.\,\cite{chen16, Zhu19, park26}. To provide context, we briefly review the key features of the phase diagram at $J/J_z = 0.6$ reported in Ref.\,\cite{park26}. The pure XXZ model exhibits magnetic ground states—Y, UUD, V, $\Psi$ and polarized\,(P)—that appear sequentially as the magnetic field \(h_z\) increases in the DMRG calculation of the ground state. Here, Y, V and $\Psi$ are the spin-supersolid phases breaking the U(1) and translational symmetries of the XXZ Hamiltonian simultaneously. These phases are further enriched when spin--orbit-induced interactions are included, which destabilize the three-sublattice order and lead to the emergence of magnetic ground states with various unit-cell sizes\,\cite{chen16, Zhu19, park26}. Notably, it was revealed that SOC drives a zero-temperature instability in these supersolids by opening a magnon gap. On the other hand, thermal fluctuations can counteract this destabilization, restoring supersolidity in the Y and $\Psi$ phases. With larger SOC-induced interactions, the system transitions into various magnetic orders \cite{park26}: two coplanar phases, namely the uniform-canted stripe (UCS) and canted stripe (CS) phases, and a non-coplanar skyrmion lattice\,(SkX) phase.

In this work, we employ the infinite density matrix renormalization group (iDMRG) method and linear spin-wave theory to investigate the spin chirality of magnetic phases with finite local moments, $\langle \mathbf{S}_j \rangle \neq 0$. Our analysis focuses specifically on the P, UCS, and CS phases. We adopt $J/J_z = 0.6$ throughout the paper as a representative value, which is the same parameter used in Ref.~\onlinecite{park26}. Remarkably, we find that a nonzero quantum expectation value $\chi_{ijk} \neq 0$ emerges not only in the non-coplanar SkX phase but also in several phases where the magnetic order is collinear or coplanar and the classical SSC vanishes ($\chi_{ijk}^{\rm cl} = \langle \Sp_i \rangle \cdot \langle \Sp_j \rangle \times \langle \Sp_k \rangle = 0$), yet the quantum average remains finite. We define this regime, where quantum correlations induce chirality absent in the classical limit, as possessing quantum scalar spin chirality (QSSC) order.

\begin{figure}[]
    \includegraphics[width=0.9\columnwidth]{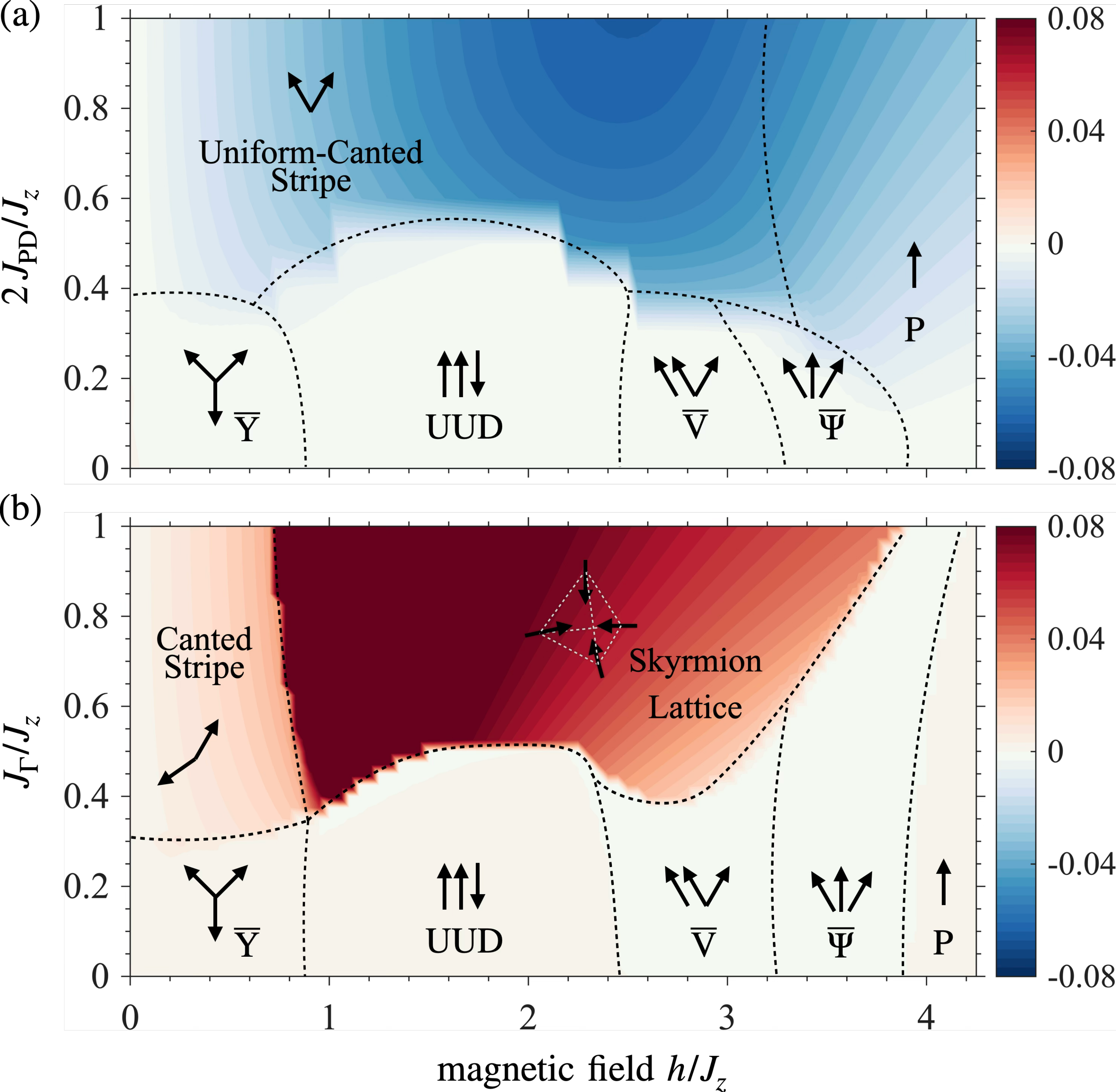}
    \caption{Phase diagram and scalar spin chirality expectation value as functions of the magnetic field $h$ and the spin-orbit-induced interactions (a) $J_{\rm PD }$, and (b) $J_{\Gamma}$, respectively. Spins in the canted stripe phase are collinear, whereas those in the uniform-canted stripe phase are coplanar.}
    \label{fig:chirality}
\end{figure}

The QSSC in the ground state is computed as
\begin{align}
    \chi &\equiv \frac{1}{N_\triangle }\sum_{\langle ijk\rangle\in\triangle}
    \left\langle \Sp_i \cdot \Sp_j \times \Sp_k \right\rangle,
    \label{eq:chi}
\end{align}
where \(\langle ijk\rangle\) denotes an anticlockwise oriented triplet 
of sites on each elementary triangular plaquette, and \(N_\triangle\) 
is the total number of such plaquettes. 
As shown in Fig.\,\ref{fig:chirality}, phases (Y, UUD, V, $\Psi$) that exist at small values of anisotropy have strongly suppressed $\chi$. Our iDMRG calculations confirm that, in these phases, the local SSC $\chi_{ijk}$ on each elementary triangle is nonzero but alternates in sign between adjacent up and down triangles, leading to nearly complete cancellation in the lattice-averaged $\chi$. This cancellation can be understood from the three-sublattice structure of these phases: for each edge shared by an up and a down triangle, the two unshared vertices carry the same sublattice index and hence the same spin expectation value, which enforces $\chi_\triangle + \chi_\triangledown = 0$ in accordance with the no-go argument of Ref.~\cite{katsura10}. In contrast, the UCS, CS, and SkX phases possess magnetic unit cells in which adjacent up and down triangles have unshared vertices belonging to different sublattices with distinct magnetic moments. The no-go condition is therefore not satisfied, and we find that all elementary triangles carry a uniform $\chi_{ijk}$ of the same sign, yielding a finite lattice-averaged $\chi$.

\begin{figure}[]
    \centering
    \includegraphics[width=0.45\textwidth]{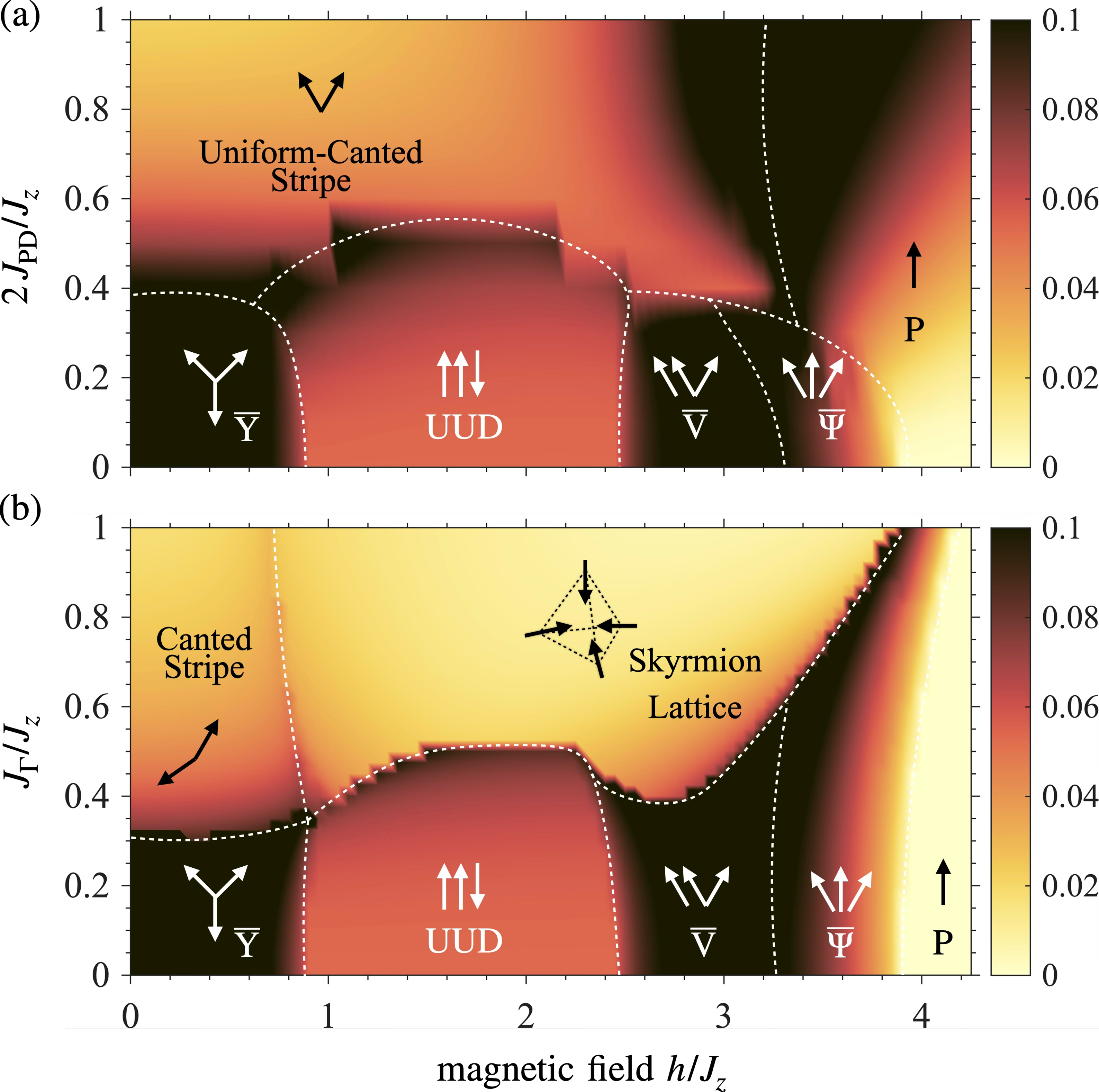}
    \caption{Spin fluctuation: $\delta S = S - N^{-1}\sum_{i} | \langle \Sp_i \rangle | $ with $S=1/2$. Regions where $\delta S\geq 0.1$ are rendered in black, corresponding to relatively large spin fluctuations.}
    \label{fig:fluctuation}
\end{figure}

\section{Magnon analysis}

As discussed in the previous section, the point group symmetry of the 
Hamiltonian is reduced from $\mathsf{D_{3d}}$ to $\mathsf{S_6}$ by the 
external magnetic field, lifting the symmetry constraint on SSC imposed 
by $C_2$ and $\sigma_d$. Furthermore, SOC breaks the 
effective time-reversal symmetry $\mathcal{T}'$ that would otherwise 
forbid a nonzero SSC even under $\mathsf{S_6}$. While these symmetry 
considerations establish that a nonzero SSC is allowed, they do not 
determine its magnitude or microscopic origin. In this section, we 
perform Holstein-Primakoff (HP) analysis for the various phases to 
demonstrate how quantum fluctuations generate a finite QSSC and to 
quantify its value.

{\it Uniform-canted Stripe phase} - As seen in the phase diagram, the uniform-canted stripe (UCS) phase appears only for $J_{\rm PD} \neq 0$. Its classical configuration is given by 
\begin{align} 
\langle \mathbf{S}_i \rangle_{\rm cl} = S (0, (-1)^{i_2}\cos \alpha, \sin \alpha ) ,
\label{eq:UCS}
\end{align}
with the lattice vector
\begin{align*} \mathbf{R}_i=i_1 \mathbf{a}_1+i_2 \mathbf{a}_2,\quad \mathbf{a}_1 =(1,0),\quad \mathbf{a}_2 = \left( -\frac{1}{2},\frac{\sqrt{3}}{2} \right),
\end{align*} 
and the classical spin magnitude $S=1/2$. The canting angle $\alpha$ obtained by minimizing the classical ground-state energy is 
\begin{align}
\sin \alpha = \frac{h_z}{J+3J_z+4J_{\rm PD}},
\label{eq:alpha0}
\end{align}
which deviates from the DMRG data of the same quantity by less than 10\%. 

\begin{figure}[t]
    \centering
    \includegraphics[width=0.48\textwidth]{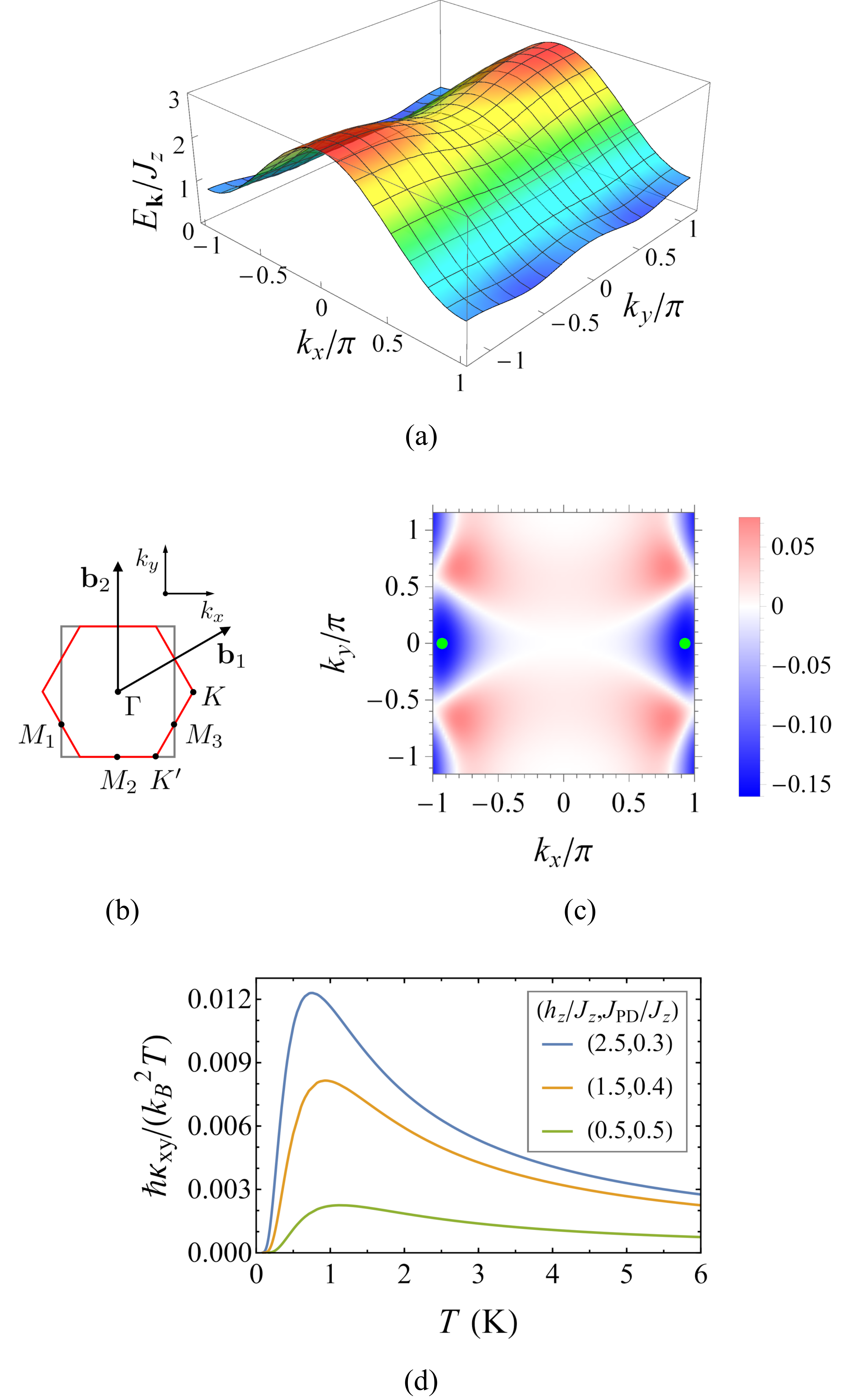}
    \caption{UCS phase of $H_{\rm XXZ} + H_{\rm PD}$: (a) magnon dispersion at $(h_z,J_{\rm PD})=(2.5,0.3)J_z$. (b) Brillouin zone (BZ) shown in red hexagon and gray rectangle as related by translation symmetry. $\mathbf{b}_1=2\pi(1,1/\sqrt{3}),\,\mathbf{b}_2=2\pi(0,2/\sqrt{3})$ are the reciprocal lattice unit vectors defined by $\mathbf{a}_n\cdot\mathbf{b}_m=2\pi\delta_{nm}$. (c) Berry curvature of the magnon band at $(h_z,J_{\rm PD})=(2.5,0.3)J_z$. The green dots mark the points of maximal magnitude in Berry curvature. (d) Thermal Hall conductivity of the magnons computed at $J_z=0.125\,\mathrm{meV}$ relevant for $\rm Na_2BaCo(PO_4)_2$~\cite{sheng25}.}
\label{fig:UCS}
\end{figure}

Proceeding with HP analysis to capture small fluctuations around the classical minimum yields a $2\times 2$ magnon Hamiltonian, 
\begin{align}
    H_{m} = \frac{1}{2}\sum_{\mathbf{k} \in {\rm BZ}} \Bigl[ \begin{pmatrix} a_\mathbf{k}^\dagger & a_{-\mathbf{k}} \end{pmatrix} \begin{pmatrix} d_\mathbf{k} & \Delta_\mathbf{k} \\ \Delta_\mathbf{k}^* & d_\mathbf{k} \end{pmatrix} \begin{pmatrix} a_\mathbf{k} \\ a_{-\mathbf{k}}^\dagger \end{pmatrix} -d_\mathbf{k} \Bigr] +E_0,
\label{eq:H_2x2}
\end{align}
with explicit expressions for $d_\mathbf{k},\,\Delta_\mathbf{k}$ and the classical ground state energy $E_0$ given in App. \ref{app:UCS-phase}. It can be diagonalized by the Bogoliubov transformation
\begin{gather}
    \begin{pmatrix} a_\mathbf{k} \\ a_{-\mathbf{k}}^\dagger \end{pmatrix} = T_\mathbf{k} \begin{pmatrix} b_\mathbf{k} \\ b_{-\mathbf{k}}^\dagger \end{pmatrix}, \nn [3pt]
    T_\mathbf{k} = \begin{pmatrix} \cosh\theta_\mathbf{k} & -e^{i\psi_\mathbf{k}}\sinh\theta_\mathbf{k} \\ -e^{-i\psi_\mathbf{k}}\sinh\theta_\mathbf{k} & \cosh\theta_\mathbf{k}\end{pmatrix} , 
    \label{eq:Bogoliubov_UCS}
\end{gather}
resulting in the quasiparticle Hamiltonian
\begin{align} 
H_m & = \sum_{{\bf k} \in {\rm BZ}} \biggl[ E_{\bf k} \Bigl( b^\dag_{\bf k} b_{\bf k}+\frac{1}{2}\Bigr) - \frac{d_\mathbf{k}}{2} \biggr] +  E_0 , \nn 
E_\mathbf{k} & = \sqrt{d_\mathbf{k}^2-|\Delta_\mathbf{k}|^2} . 
\label{eq:UCS_dispersion}
\end{align} 
The rotation angles $\theta_\mathbf{k}$, $\psi_\mathbf{k}$ are defined by 
\begin{align*} 
\tanh{2\theta_\mathbf{k}} = |\Delta_\mathbf{k}|/d_\mathbf{k},  ~~ 
\Delta_\mathbf{k} =|\Delta_\mathbf{k}|e^{i\psi_\mathbf{k}}, 
\end{align*} 
respectively. The magnon spectrum \eqref{eq:UCS_dispersion} exhibits the excitation gap
\begin{align}
    \Delta = \sqrt{2J_{\rm PD}(J-J_z+4J_{\rm PD})}\cos\alpha
\end{align}
at the high-symmetry points $M_1$ and $M_3$, and the maximum at $M_2$. The magnon dispersion along with its Brillouin zone is depicted in Fig. \ref{fig:UCS}. 

The magnitude of spin fluctuation in the magnon theory is 
\begin{align}
    \delta S & = S- N^{-1} \sum_j |\langle \Sp_j \rangle | = N^{-1} \sum_{\mathbf{k}\in{\rm BZ}}  \langle a_\mathbf{k}^\dagger a_\mathbf{k}\rangle  \nn 
    & = \frac{1}{N}\sum_{\mathbf{k}\in{\rm BZ}}  \left[ \sinh^2\theta_\mathbf{k} + n_\mathbf{k}(T)\cosh{2\theta_\mathbf{k}} \right] ,
\label{eq:deltaS-UCS}
\end{align}
where $N$ is the number of sites, and $n_\mathbf{k}(T)=\lbrace\exp{[E_\mathbf{k}/\kb T]}-1\rbrace^{-1}$ is the Bose-Einstein distribution. When not too close to the phase boundary, iDMRG gives typical values of $\delta S$ about 10\% of the classical value $S=1/2$. The HP estimates for $\delta S$ are smaller than those of DMRG; for instance, we get $\delta S_{\rm DMRG}=0.0568$ and $\delta S_{\rm HP} = 0.0340$ at $(h_z/J_z, J_{\rm PD}/J_z) = (2.5, 0.3)$. (A more complete comparison between DMRG and HP calculations is given in App.~\ref{app:UCS-phase}.) Overall, $\delta S/S$ remains sufficiently small to justify the use of HP theory as long as $k_BT \lesssim \Delta$.

The QSSC defined in Eq.~\eqref{eq:chi} can be written as 
\begin{align}
    \chi = \frac{1}{2N}\sum_i \langle \Sp_{\mathbf{R}_i} \cdot \Sp_{\mathbf{R}_i+\mathbf{a}_2} \times \Sp_{\mathbf{R}_i-\mathbf{a}_1} + \Sp_{\mathbf{R}_i} \cdot \Sp_{\mathbf{R}_i-\mathbf{a}_2} \times \Sp_{\mathbf{R}_i+\mathbf{a}_1}\rangle
    \label{eq:chi_i}
\end{align}
involving the sum over both up and down triangles. The classical average of $\chi$ given by, \textit{e.g.} $\langle \Sp_{\mathbf{R}_i} \rangle \cdot \langle \Sp_{\mathbf{R}_i+\mathbf{a}_2}\rangle \times \langle \Sp_{\mathbf{R}_i-\mathbf{a}_1} \rangle$ is strictly zero in the UCS phase since the spins are coplanar, but due to quantum fluctuations, the following average becomes nonzero: 
\begin{align} 
& \langle \Sp_{\mathbf{R}_i} \cdot \Sp_{\mathbf{R}_i+\mathbf{a}_2} \times \Sp_{\mathbf{R}_i-\mathbf{a}_1} \rangle  \nn 
&= \langle \Sp_{\mathbf{R}_i} \rangle_{\rm cl} \cdot \langle \delta \Sp_{\mathbf{R}_i+\mathbf{a}_2} \times \delta \Sp_{\mathbf{R}_i-\mathbf{a}_1} \rangle + {\rm cyclic\,} {\rm permutations},
\label{eq:chi_vsc}
\end{align}
when expanded to the second order in $\delta \Sp_i = \Sp_i - \langle \Sp_i \rangle_{\rm cl}$. The QSSC is now given as the product of the magnetic order $\langle \Sp_i \rangle_{\rm cl}$ and the VSC $\langle \delta \Sp_j \times \delta \Sp_k \rangle$. The VSC points in the orthogonal direction to the magnetization at the classical level and acquires a parallel component through quantum corrections, which is captured by the magnon theory. 
One can show that only the $y$-component of the VSC contributes to $\chi$~\cite{katsura10}:   
\begin{align}
&\langle\delta\mathbf{S}_{\mathbf{R}_i}\times\delta\mathbf{S}_{\mathbf{R}_i+\mathbf{a}_2}\rangle^y = \frac{(-1)^{i_2}}{2N}\cos\alpha \times \nn
    & \quad \sum_{\mathbf{k}\in{\rm BZ}} (2n_\mathbf{k}(T)+1)\sin\psi_\mathbf{k} \sinh{2\theta_\mathbf{k}}\cos(\mathbf{k}\cdot\mathbf{a}_2).
    \label{eq:VSC_UCS}
\end{align}
Together with $\langle S^y_i \rangle = S (-1)^{i_2} \cos \alpha$, the QSSC $\chi$ becomes 
\begin{align}
 \chi = \frac{\cos^2\alpha}{2N} \sum_{\mathbf{k}\in{\rm BZ}} (2n_\mathbf{k}(T)+1)\sin\psi_\mathbf{k} \sinh{2\theta_\mathbf{k}}\cos(\mathbf{k}\cdot\mathbf{a}_2).
\label{eq:chi-UCS}
\end{align}
This detailed derivation of the magnon formula for $\chi$ demonstrates that, in the UCS phase, the leading contribution to the QSSC arises from the quantum-fluctuation-induced VSC parallel to the magnetization direction.

\begin{figure*}[t]
    \centering
    \includegraphics[width=\textwidth]{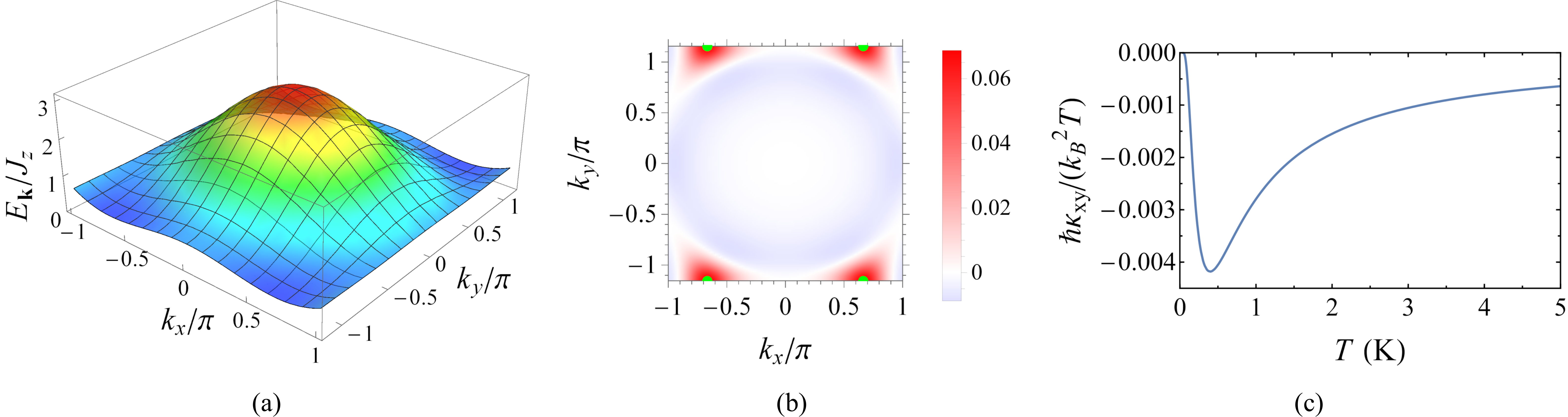}
    \caption{Polarized phase in $H_{\rm XXZ} + H_{\rm PD}$: (a) magnon dispersion, (b) Berry curvature, and (c) thermal Hall conductivity at $(h_z,J_{\rm PD})=(4.25,0.05)J_z,\,J_z=0.125\,\mathrm{meV}$. The BZ is the same as in the UCS phase. The green dots in (b) mark the points of maximal Berry curvature which coincide with the positions of the excitation gap $\Delta$.}
\label{fig:polarized}
\end{figure*}

The HP analysis gives $\chi$ of the same sign as the iDMRG but about 10--40\% smaller in magnitude. Specifically, $(h_z,J_{\rm PD})=(2.5,0.3)J_z$ yields $\chi_{\rm HP}=-0.0485$ and $\chi_{\rm DMRG}=-0.0548$ (see App.~\ref{app:UCS-phase} for a detailed comparison.) Further corrections to the quadratic magnon model are discussed in App.~\ref{app:UCS-phase}. Most importantly, the nonzero QSSC $\chi \neq 0$ arises only for $J_{\rm PD} \neq 0$, verifying that the source of QSSC is indeed the SOC.

A nonzero QSSC implies magnon thermal Hall effect (THE)~\cite{katsura10}. We compute the thermal Hall conductivity $\kappa_{xy}$~\cite{matsumoto14} in HP theory,
\begin{align}
    \kappa_{xy}=-\frac{k_B^2T}{\hbar A} \sum_{\mathbf{k}\in{\rm BZ}} \left[c_2(n_\mathbf{k})-\frac{\pi^2}{3}\right]\Omega_\mathbf{k},
\end{align}
where $c_2(x)=\int_0^x ds [\ln(1+s^{-1})]^2$ and $A=(\sqrt{3}/2)N$ is the area of the system obtained by the size of the unit cell times its number. The Berry curvature is defined as~\cite{matsumoto13}
\begin{align}
    \Omega_\mathbf{k}=i\epsilon_{\mu\nu}\left[\sigma_3\frac{\partial T_\mathbf{k}^\dagger}{\partial k_\mu}\sigma_3\frac{\partial T_\mathbf{k}}{\partial k_\nu}\right]_{11}=\epsilon_{\mu\nu}\frac{\partial\psi_\mathbf{k}}{\partial k_\mu}\frac{\partial\theta_\mathbf{k}}{\partial k_\nu}\sinh{2\theta_\mathbf{k}},
\label{eq:Berry_UCS}
\end{align}
with $T_\mathbf{k}$ the Bogoliubov transformation matrix in Eq. \eqref{eq:Bogoliubov_UCS} and $\mu, \nu = x,y$. The Chern number of the magnon band is zero, consistent with the general conclusion that the sum of Chern numbers over all bands must vanish~\cite{matsumoto13}. The calculated thermal Hall conductivity of the UCS phase, shown in Fig. \ref{fig:UCS}(d), exhibits activated behavior $\kappa_{xy}\propto e^{-\Delta/(k_BT)}$ at $T \ll \Delta$, saturates to a constant value $\kappa_{xy} \rightarrow \frac{k_B}{\hbar A}\sum_{\mathbf{k}\in{\rm BZ}} E_{\mathbf{k}}\Omega_\mathbf{k}$ at $T\gg \Delta$, and develops a peak at $\kb T\simeq \Delta$. At all temperatures, $\kappa_{xy}$ remains positive. This can be traced back to the fact that the Berry curvature is negative on average near the excitation gap, where the dominant contribution to $\kappa_{xy}$ arises, and reaches its minimum at the points $(k_x,k_y)=(\pm a\pi,0)$ with $a\lesssim 1$. The factor $a$ depends on the position in the phase diagram and takes the value $0.93$ at $(h_z,J_{\rm PD})=(2.5,0.3)$ (See Fig. \ref{fig:UCS}(c)).
\\

{\it Polarized phase} - The polarized (P) phase occurs in the phase diagram of $H_{\rm XXZ} + H_{\rm PD}$ under a higher magnetic field $h_z$ compared to the UCS phase. 
According to Eq.~\eqref{eq:alpha0}, the transition to the fully-polarized phase occurs at $h_z = J+3J_z+4J_{\rm PD}$ as the $y$-component of the magnetization goes smoothly to zero in the manner of a second-order phase transition. 

While the no-go argument~\cite{katsura10} predicts zero QSSC at the quadratic level in the P phase, the actual iDMRG results as shown in Fig.~\ref{fig:chirality}(a) give $\chi$ that varies smoothly across the phase boundary and remains nonzero in the P phase. The magnon Hamiltonian in the P phase takes the same form as in the UCS phase (Eqs. \eqref{eq:H_2x2} to \eqref{eq:UCS_dispersion}), with explicit expressions for the matrix elements $d_\mathbf{k},\,\Delta_\mathbf{k}$ and the classical ground-state energy $E_0$ shown in App.~\ref{app:P-phase}. The excitation gap at the $K$-point is $\Delta=h_z-\frac{3}{2}J-3J_z$. The magnon band and its Berry curvature are plotted in Fig. \ref{fig:polarized}. The QSSC $\chi$ is indeed zero at quadratic order in magnon theory, consistent with the no-go argument. 

Quite interestingly, QSSC computed to fourth-order in magnon expansion gives a nonzero expression 
\begin{align}
    \chi & = \frac{1}{2N}\sum_i \langle \delta\Sp_{\mathbf{R}_i} \cdot \delta\Sp_{\mathbf{R}_i+\mathbf{a}_2} \times \delta\Sp_{\mathbf{R}_i-\mathbf{a}_1} \nn
    &~~~~~~~~~~~~~~ + \delta\Sp_{\mathbf{R}_i} \cdot \delta\Sp_{\mathbf{R}_i-\mathbf{a}_2} \times \delta\Sp_{\mathbf{R}_i+\mathbf{a}_1}\rangle \nn
[6pt]
    & = -\frac{1}{8N^2} \sum_{\mathbf{k},\mathbf{p}} \begin{vmatrix}
        \cos{(\mathbf{k}\cdot\mathbf{a}_1)} & \cos{(\mathbf{k}\cdot\mathbf{a}_2)} & \cos{(\mathbf{k}\cdot\mathbf{a}_3)} \\
        \cos{(\mathbf{p}\cdot\mathbf{a}_1)} & \cos{(\mathbf{p}\cdot\mathbf{a}_2)} & \cos{(\mathbf{p}\cdot\mathbf{a}_3)} \\
        1 & 1 & 1
    \end{vmatrix} \nn 
    & ~~~~~~~~~~~~~~ \times \sin{(\psi_\mathbf{k}-\psi_\mathbf{p})}\sinh{2\theta_\mathbf{k}}\sinh{2\theta_\mathbf{p}}.
\label{eq:chi4_UCS}
\end{align}
See App.~\ref{app:P-phase} for a detailed derivation of the formula. Numerical evaluation of the formula gives $\chi_{\rm HP} =-0.0033$ at $(h_z , J_{\rm PD}) =( 4.25 ,\,0.05)J_z$, which is of the same sign as the DMRG result $\chi_{\rm DMRG} = -0.0009$ at the same parameter value. The development of QSSC is again accompanied by a finite thermal Hall response $\kappa_{xy}$, as shown in Fig. \ref{fig:polarized}(c). The sign reversal compared to $\kappa_{xy}$ in the UCS phase is due to the dominant Berry curvature being of opposite signs in the UCS and P phases. The appearance of nonzero QSSC in the iDMRG calculation is beyond the no-go argument and is captured by pushing the magnon analysis to higher order.  

There is another P phase in the phase diagram of $H_{\rm XXZ} + 
H_\Gamma$ as shown in Fig.~\ref{fig:chirality}(b), which shows no 
discernible QSSC. This can be understood by noting that the fully 
polarized state $|\!\uparrow\uparrow
\cdots\rangle$ is an exact eigenstate of $H_\Gamma$ with eigenvalue 
zero. To see this, observe that $H_\Gamma$ acting on 
the fully polarized state produces single spin-flip excitations 
$S_k^-|\!\uparrow\uparrow
\cdots\rangle$ at each site $k$, whose amplitude is 
proportional to $\sum_{\langle ij \rangle} e^{i\varphi_{ij}} = 0$,
where $\varphi_{ij} = 0,\,\pm 2\pi/3$ are the bond-dependent phases of the 
$\Gamma$ interaction on the three nearest-neighbor bond directions\,[Eq.\,\eqref{eq:hamiltonian}]. 
The cancellation ensures $H_\Gamma|\!\uparrow\uparrow
\cdots\rangle = 0$ exactly. 
Consequently, whenever the fully polarized state is the ground 
state of $H_{\rm XXZ}$, it remains the exact ground state of 
$H_{\rm XXZ}+H_\Gamma$ as a product state with identically vanishing 
quantum fluctuations and QSSC. This stands in contrast to the PD 
interaction, which generates two-magnon pair excitations 
$S_i^- S_j^-|\!\uparrow\uparrow
\cdots\rangle$ on each bond. These pair amplitudes do 
not sum over the three bond directions at a single site and therefore do 
not cancel, resulting in finite quantum fluctuations and a nonzero QSSC 
in the P phase of $H_{\rm XXZ}+H_{\rm PD}$.
\\

\begin{figure}[h]
    \centering
    \includegraphics[width=\columnwidth]{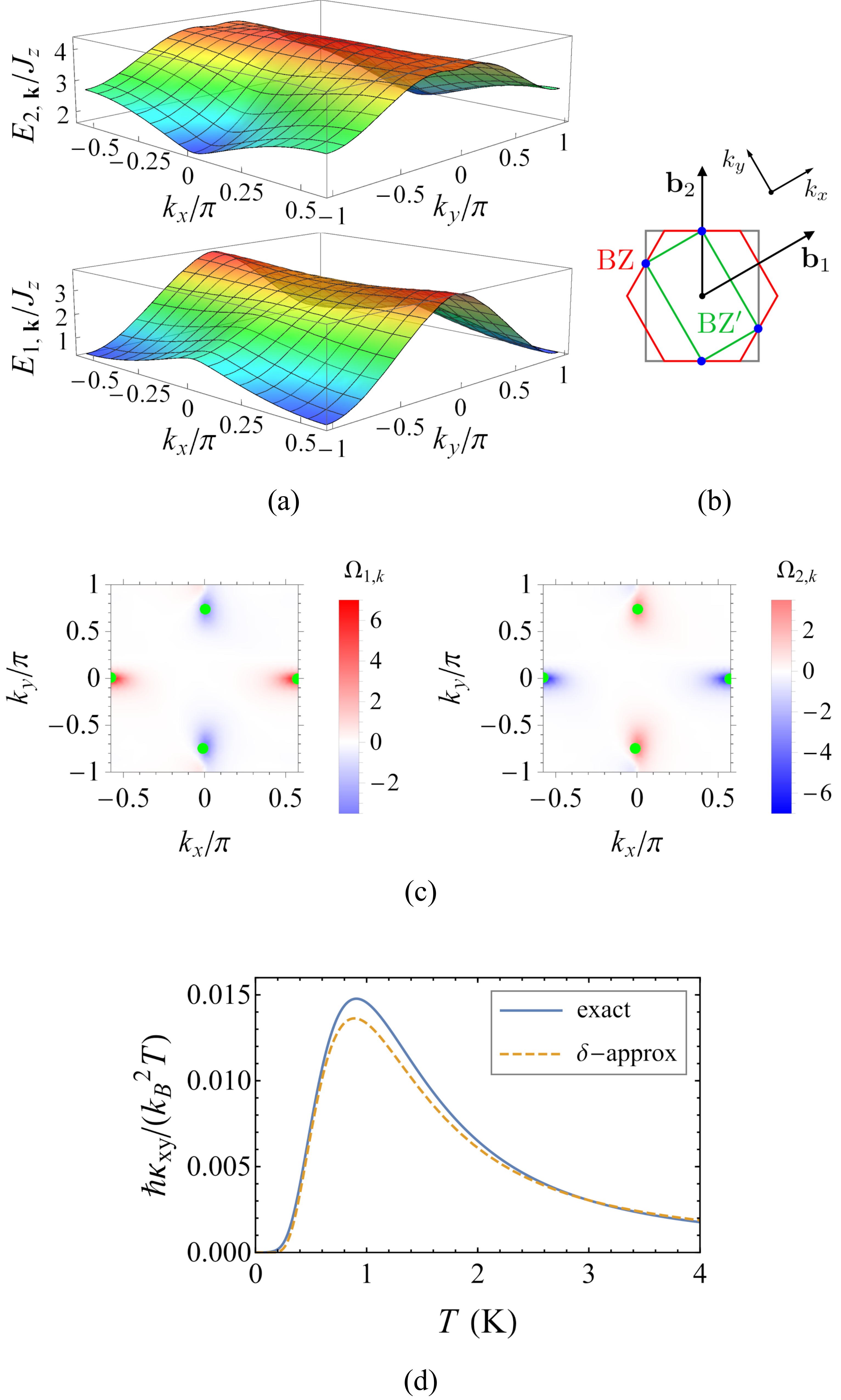}
    \caption{CS phase in $H_{\rm XXZ} + H_\Gamma$: (a) Two magnon bands $(E_{1,{\bf k}}, E_{2,{\bf k}})$. (b) Brillouin zone (BZ) and the reduced Brillouin zone (${\rm BZ}'$). The blue dots indicate the position of the excitation gap. (c) Berry curvature of the two magnon bands. The green dots mark the points with the maximal and minimal Berry curvature and nearly coincide with the points of minimum energy separation $E_{2,{\bf k}}-E_{1,{\bf k}}$. (d) Thermal Hall conductivity $\kappa_{xy}$. The blue solid line indicates the prediction of the exact magnon theory, and the orange dashed line was obtained by approximating the Berry curvature by delta functions. All panels are calculated using $(J_\Gamma,h_z)=(1.0,0.2)J_z$ with $J_z=0.125\,{\rm meV}$.}
\label{fig:CS}
\end{figure}

{\it Canted stripe phase} - 
The canted stripe (CS) phase occurs in the phase diagram for 
$H_{\XXZ} + H_\Gamma$ and develops QSSC despite having zero $\chi$ 
with respect to the classical spin configuration, which is given by
\begin{align}
    \langle \mathbf{S}_i \rangle_{\rm cl} 
    &= \eta_i S
    \Big(
      \sin\Theta_i \cos\phi,\;
      \sin\Theta_i \sin\phi,\;
      \cos\Theta_i
    \Big) ,
    \label{eq:CS}
\end{align}
with $\eta_i = (-1)^{i_1},\,S=1/2$ and
\begin{align}
    \Theta_i 
    = \theta - \frac{1+\eta_i}{2}\psi_A + \frac{1-\eta_i}{2}\psi_B .
\label{eq:theta_i}
\end{align}
The sublattice-dependent polar angles are given by
\begin{align*}
    \Theta_A = \theta - \psi_A ,\,\, \Theta_B = \theta + \psi_B ,
\end{align*}
where $A$ ($B$) sublattice denotes sites with $\eta_i = +1$ $(-1)$. There are altogether four angles $(\phi, \theta, \psi_A, \psi_B )$ which characterize the ground state classically, and can be obtained by minimizing the classical energy in the manner shown in App.~\ref{app:CS-phase}. In particular, the angle $\phi = \pi/6$ obtained from energy minimization is in excellent agreement with the DMRG result. At $h_z=0$, the classical spin configuration becomes collinear, $\psi_A=\psi_B$, and the angle $\theta$ is obtained classically as 
\begin{align}
    \tan 2\theta
    = \,\frac{4J_\Gamma}{J_z - J}.
\label{eq:theta_h=0}
\end{align}
For $(J,J_\Gamma)=(0.6,1.0)J_z$ we obtain
$\theta \simeq 42.14^\circ$, which lies very close to the DMRG value $\theta_{\rm DMRG} \simeq 43.04^\circ$. As we switch on the magnetic field $h_z$, the two angles $\psi_A, \psi_B$ increase by the sublattice-dependent amount, which is also the case in DMRG. 

To develop the magnon theory of fluctuations, we introduce the Nambu spinor
\begin{align*}
    \Psi_{\mathbf{k}}
    =
    \begin{pmatrix}
        a_{\mathbf{k}} & 
        b_{\mathbf{k}} & 
        a_{-\mathbf{k}}^\dagger &
        b_{-\mathbf{k}}^\dagger
    \end{pmatrix}^\top ,
\end{align*}
and write the magnon Hamiltonian as
\begin{gather}
    H_m
     =
    \sum_{\mathbf{k}\in {\rm BZ}'}
    \Bigl[
        \frac{1}{2}\,
        \Psi_{\mathbf{k}}^\dagger
        H_{\mathbf{k}}
        \Psi_{\mathbf{k}}
        -
        \frac{1}{2}(\epsilon_A+\epsilon_B)
    \Bigr]
    + E_0, \nn 
H_{\mathbf{k}}
     =
    \begin{pmatrix}
        h_{\mathbf{k}} & \Delta_{\mathbf{k}} \\
        \Delta^\dagger_{\mathbf{k}} & h^\top_{-\mathbf{k}}
    \end{pmatrix}.
\label{eq:H_CS}
\end{gather}
Derivations of the magnon model together with explicit expressions for the $2\times2$ matrices $h_\mathbf{k},\,\Delta_\mathbf{k}$ and the constants $\epsilon_A,\epsilon_B, E_0$ are given in App. \ref{app:CS-phase}. The reduced Brillouin zone (${\rm BZ}'$) is shown schematically in Fig.~\ref{fig:CS}(b). 

Transformation to the quasiparticle basis takes place as 
\begin{align}
    \Psi_{\mathbf{k}}
    =
    T_{\mathbf{k}}\,
    \Phi_{\mathbf{k}},
\qquad
    \Phi_{\mathbf{k}}
    =
    \begin{pmatrix}
        \alpha_{\mathbf{k}} & 
        \beta_{\mathbf{k}} &
        \alpha_{-\mathbf{k}}^\dagger &
        \beta_{-\mathbf{k}}^\dagger
    \end{pmatrix}^\top ,
\label{eq:CS-Bogol}
\end{align}
where the Bogoliubov matrix $T_{\mathbf{k}}$ is defined by 
\begin{gather*}
    T_{\mathbf{k}}^\dagger\Sigma_3 T_{\mathbf{k}}=\Sigma_3,\quad \Sigma_3 = \begin{pmatrix}
        \mathbb{I}_2 & 0 \\
        0 & -\mathbb{I}_2
    \end{pmatrix}, \nn 
    \Sigma_3 H_{\mathbf{k}}T_{\mathbf{k}} =
    T_{\mathbf{k}}\,
    \mathrm{diag}\!\left(
        E_{1,\mathbf{k}},\,
        E_{2,\mathbf{k}},\,
        -E_{1,-\mathbf{k}},\,
        -E_{2,-\mathbf{k}}
    \right) .
\end{gather*}
Using this, the quasiparticle Hamiltonian becomes
\begin{align}
    H_m
    &=
    \sum_{\mathbf{k}\in{\rm BZ}'}
    \left[E_{1,\mathbf{k}}\, \alpha_{\mathbf{k}}^\dagger \alpha_{\mathbf{k}}
    + E_{2,\mathbf{k}}\, \beta_{\mathbf{k}}^\dagger \beta_{\mathbf{k}}\right]
    + E_2 ,
    \label{eq:Hm_diag}
\end{align}
with
\begin{align*}
    E_2
    =
    E_0
    +
    \sum_{\mathbf{k}\in{\rm BZ}'}
    \left[
        \frac{1}{2}\sum_{\sigma=1}^{2} E_{\sigma,\mathbf{k}}
        - \frac{1}{2}(\epsilon_A+\epsilon_B)
    \right] . 
\end{align*}
The two magnon energy bands are shown in Fig. \ref{fig:CS}(a). The energy gap, \textit{i.e.} the lowest energy of $E_{1,{\bf k}}$, occurs at the corners of ${\rm BZ}'$, ${\bf k} = (\pm 1/\sqrt{3},\pm 1)\pi$, as indicated in Fig.~\ref{fig:CS}(b). We have verified that $\delta S$ in the free-magnon theory is sufficiently small compared to $S=1/2$ to justify its use.

As in the UCS phase, a nonzero QSSC in the CS phase is allowed by the no-go argument~\cite{katsura10} and can be related to a component of VSC emerging from quantum fluctuations. A step-by-step derivation, given in App.~\ref{app:CS-phase}, yields
\begin{align}
\chi & =
    \frac{i\big(
        1 \!+\! \cos(\psi_A + \psi_B)
    \big)}{8N}\sum_{\mathbf k\in {\rm BZ}'}
    \Big[
        A_{\mathbf k}\, \langle a_\mathbf{k} b_{-\mathbf{k}} \rangle
        - {\rm c.c.} 
    \Big], \nn 
A_{\bf k} & = -1-e^{2i\mathbf{k}\cdot\mathbf{a}_1}+e^{-i\mathbf{k}\cdot\mathbf{a}_2}+e^{i\mathbf{k}\cdot(2\mathbf{a}_1+\mathbf{a}_2)} .
\label{eq:chi_CS}
\end{align}
The formula for $\chi$ follows once again from the product of the local magnetization times the VSC along that direction. For $(h_z,J_\Gamma)=(0.2,1.0)J_z$, we obtain at zero temperature $\chi_{\rm HP} \simeq 0.0057$. This value captures the correct sign and magnitude of the DMRG result $\chi_{\mathrm{DMRG}} \simeq 0.0063$ up to a relative difference of 10\%.

The Berry curvature for each band is computed as
\begin{align}
    \Omega_{\sigma,\mathbf{k}}= i\epsilon_{\mu\nu}\!\left[\Sigma_3(\partial_{k_\mu}T_{\mathbf{k}}^\dagger)\Sigma_3
(\partial_{k_\nu}T_{\mathbf{k}})\right]_{\sigma\sigma} .
    \label{eq:Omega_uonly}
\end{align}
and shown in Fig.~\ref{fig:CS}(c). The Chern number of each band vanishes. The magnon thermal Hall conductivity is obtained by summing the contributions from the two magnon bands,
\begin{align}
\kappa_{xy}(T) = 
-\frac{k_B^2 T}{\hbar A}
\sum_{\mathbf{k}\in{\rm BZ}'}
\sum_{\sigma=1}^{2}
\left[
c_2\!\left(n_{\sigma,{\bf k}}(T)\right)-\frac{\pi^2}{3}
\right]
\Omega_{\sigma, \mathbf{k}}, 
\end{align}
where $n_{\sigma,{\bf k}}(T)=\lbrace\exp{[E_{\sigma,\mathbf{k}}/\kb T]}-1\rbrace^{-1}$. As seen in Fig.~\ref{fig:CS}(d), it is comparable to $\kappa_{xy}$ found in other phases, despite the very small magnitude of $\chi$. Notice that the Berry curvature in the CS phase is highly concentrated and reaches enormous strengths around the anti-crossing points of the two bands. Approximating the Berry curvature distribution as delta functions faithfully captures the $\kappa_{xy}$ plot. For the other two phases, UCS and P, there is only one magnon band and the Berry curvature distribution is much more diffuse. 

\section{Summary and discussion}
We have investigated the phase diagram of the quantum spin-1/2 XXZ 
Hamiltonian on a triangular lattice supplemented by spin-orbit-induced 
exchange interactions. The model is known to support several types of 
spin-supersolid phases as well as the skyrmion crystal phase. Here we 
uncover hitherto unknown phases of this model supporting the quantum 
scalar spin chirality (QSSC). Unlike its classical counterpart arising 
solely from the non-coplanar orientation of the magnetic moments such 
as found in the skyrmion crystal phase, this quantum order arises 
{\it despite} having coplanar or collinear arrangement of magnetic 
moments.

The symmetry analysis reveals that two conditions must be simultaneously 
met for a nonzero SSC to be allowed: (i) the external magnetic field 
reduces the point group from $\mathsf{D_{3d}}$ to $\mathsf{S_6}$, 
removing the $C_2$ and $\sigma_d$ operations that would otherwise 
forbid SSC, and (ii) spin-orbit coupling breaks the effective 
time-reversal symmetry $\mathcal{T}'=U_y(\pi)\mathcal{T}$ that 
prohibits SSC even under $\mathsf{S_6}$. Once these symmetry 
constraints are lifted, quantum fluctuations generate a finite QSSC even though the classical spin configuration carries no chirality. In this work, we have identified several phases in which such phenomenon takes place.

The Holstein-Primakoff analysis clarifies the microscopic mechanism 
underlying the QSSC in each phase. In the UCS and CS phases, the QSSC 
arises already at quadratic order in the magnon expansion through the emergence of a vector spin chirality component parallel to the local magnetization direction, 
which is absent in the classical spin configuration. In the polarized 
phase of $H_{\rm XXZ}+H_{\rm PD}$, the magnon contribution vanishes at the quadratic order in accord with the no-go argument of Ref.~\cite{katsura10}, but a 
finite QSSC still emerges at fourth order in the magnon expansion. 
Remarkably, in the polarized phase of $H_{\rm XXZ}+H_\Gamma$, 
the fully polarized state is an exact eigenstate of $H_\Gamma$ due to 
the $C_3$-symmetric cancellation of single-magnon amplitudes on the 
triangular lattice, resulting in identically vanishing quantum 
fluctuations and QSSC. In all phases where HP theory is applicable, 
the analytical results agree with the iDMRG data in sign and capture 
the magnitude to within 10--40\%. 

The finite QSSC is accompanied by nontrivial magnon Berry curvature 
and a magnon thermal Hall effect. We have computed the thermal Hall 
conductivity $\kappa_{xy}$ for the UCS, polarized, and CS phases, 
finding that $\kappa_{xy}$ exhibits activated behavior at low 
temperatures and a sign that reflects the dominant Berry curvature 
near the magnon gap. The magnon thermal Hall effect provides a direct experimental 
signature of the QSSC. Since spin-orbit-coupled triangular-lattice 
antiferromagnets with exchange energy scales of order 
$\sim 0.1\,\mathrm{meV}$ would place the relevant temperature 
scale in the sub-kelvin regime, triangular-lattice antiferromagnets may serve 
as promising platforms for observing the thermal Hall signatures 
predicted in this work.

\acknowledgments JHH was supported by the National Research Foundation of Korea (NRF) grant funded by the Korea government(MSIT) (
No. RS-2024-00410027).
H.-Y.L was supported by the Basic Science Research Program through the National Research Foundation of Korea funded by the Ministry of Science and ICT [Grant No. RS-2023-00220471, RS-2025-16064392].
S.P. was supported by the National Research Foundation of Korea (NRF) grant funded by the Korea government (MSIT) (Grant No. RS-2025-00559286); 
by the Nano \& Material Technology Development Program through the NRF funded by MSIT (Grant Nos. RS-2024-00451261 and RS-2023-00281839); 
by the National Measurement Standards Services and Technical Support for Industries funded by the Korea Research Institute of Standards and Science (KRISS) (KRISS-2025-GP2025-0015);
and by the Global Partnership Program of Leading Universities in Quantum Science and Technology through the NRF funded by MSIT (Grant Nos. RS-2025-08542968 and RS-2023-00256050).

\bibliography{ref}

\newpage

\begin{widetext}
\appendix
\section{HP analysis of the uniform-canted stripe phase}
\label{app:UCS-phase}

The ground-state degeneracy of the uniform-canted stripe (UCS) phase is threefold, with the classical spin configuration
\begin{align}
    \langle \Sp_{i} \rangle_{\rm cl} = S(\eta_i\cos\alpha\cos\phi,\eta_i\cos\alpha\sin\phi,\sin\alpha),
\label{eq:UCS_general}
\end{align}
where $\eta_i=(-1)^{i_1},\,(-1)^{i_2}$, or $(-1)^{i_1+i_2}$. The corresponding azimuthal angles $\phi= \pi/6,\,\pi/2,-\pi/6$ are obtained by substituting Eq. \eqref{eq:UCS_general} into the Hamiltonian \eqref{eq:hamiltonian} and minimizing the classical energy with respect to $\phi$. In the following, we fix $\eta_i=(-1)^{i_2}$ and $\phi=\pi/2$.

As in the standard HP procedure, the local reference frame at each site is rotated such that the local spin expectation values \eqref{eq:UCS} are aligned with the new local z-axes. This corresponds to the canonical transformation of the spin operators,
\begin{align*}
    S_i^x = \tilde{S}_i^x , \quad
    S_i^y = \tilde{S}_i^y \sin \alpha +(-1)^{i_2} \tilde{S}_i^z \cos \alpha , \quad
    S_i^z = -(-1)^{i_2} \tilde{S}_i^y \cos \alpha + \tilde{S}_i^z \sin \alpha.
\end{align*}
The rotated spin operators are then expressed in terms of bosonic creation and annihilation operators via the HP transformation,
\begin{align*}
    \tilde{S}_i^+ = \tilde{S}_i^x+i\tilde{S}_i^y = a_i, \quad
    \tilde{S}_i^- = \tilde{S}_i^x-i\tilde{S}_i^y = a_i^\dagger, \quad
    \tilde{S}_i^z = \frac{1}{2}-a_i^\dagger a_i.
\end{align*}
Although the operator square root $\sqrt{1-a_i^\dagger a_i}$ in the usual HP representation is approximated by unity and not all eigenstates of $a_i^\dagger a_i$ correspond to physical spin eigenstates, one can show that for spin-$1/2$ without on-site interactions the above form of the HP transformation leaves the Hamiltonian exact and the energy eigenspace physical~\cite{vogl20,hucht21}.

Since the spin fluctuation $\delta S = \frac{1}{N}\sum_i \langle a_i^\dagger a_i\rangle$ is small within the UCS phase away from the phase boundaries, terms of higher than second order in the ladder operators are expected to be negligible. The leading contributions to the Hamiltonian are
\begin{align}
    E_0 =& N \left( -\frac{J}{4}\cos^2\alpha+\frac{3}{4}J_z\sin^2\alpha-J_{\rm PD}\cos^2\alpha-\frac{h_z}{2}\sin\alpha\right),
    \nn
    H_1 =& i\sqrt{N}\left[\left(\frac{J}{4}+\frac{3}{4}J_z+J_{\rm PD}\right)\sin 2\alpha-\frac{h_z}{2}\cos\alpha\right](a_{\mathbf{b}_2/2}-a_{\mathbf{b}_2/2}^\dagger), 
    \nn
    H_2 =& \frac{1}2\sum_{\mathbf{k}\in{\rm BZ}} \left[\begin{pmatrix} a_\mathbf{k}^\dagger & a_{-\mathbf{k}} \end{pmatrix} \begin{pmatrix} d_\mathbf{k} & \Delta_\mathbf{k} \\ \Delta_\mathbf{k}^* & d_\mathbf{k} \end{pmatrix} \begin{pmatrix} a_\mathbf{k} \\ a_{-\mathbf{k}}^\dagger \end{pmatrix} - d_\mathbf{k}\right] = \sum_{\mathbf{k}\in{\rm BZ}} E_\mathbf{k} \left(b_\mathbf{k}^\dagger b_\mathbf{k} + \frac{1}{2}\right) + 2E_0+N\frac{h_z}{2}\sin\alpha \equiv \sum_{\mathbf{k}\in{\rm BZ}} E_\mathbf{k} b_\mathbf{k}^\dagger b_\mathbf{k} + E_2,
    \nn
    d_\mathbf{k} =& \frac{J}{2}(1+\sin^2\alpha)\sum_{n=1}^3\cos(\mathbf{k}\cdot\mathbf{a}_n)+\frac{J_z}{2}\cos^2\alpha\sum_{n=1}^3 P_n\cos(\mathbf{k}\cdot\mathbf{a}_n)+J_{\rm PD}\cos^2\alpha\sum_{n=1}^3 C_n\cos(\mathbf{k}\cdot\mathbf{a}_n)-\frac{4E_0}{N}-h_z\sin\alpha, \nn
    \Delta_\mathbf{k} =& \frac{J}{2}\cos^2\alpha\sum_{n=1}^3\cos(\mathbf{k}\cdot\mathbf{a}_n)-\frac{J_z}{2}\cos^2\alpha\sum_{n=1}^3 P_n\cos(\mathbf{k}\cdot\mathbf{a}_n)+J_{\rm PD}(1+\sin^2\alpha)\sum_{n=1}^ 3C_n\cos(\mathbf{k}\cdot\mathbf{a}_n) \nn
    &-i\sqrt{3}J_{\rm PD}\sin\alpha [\cos(\mathbf{k}\cdot\mathbf{a}_2)-\cos(\mathbf{k}\cdot\mathbf{a}_3)],
\label{eq:H_full_UCS}
\end{align}
where $\mathbf{a}_3=-\mathbf{a}_1-\mathbf{a}_2=-(1,\sqrt{3})/2$ and the coefficients $P_n$, $C_n$ are defined by $P_1=C_1=1$, $P_2=P_3=-1$, $C_2=C_3=-1/2$.

To reflect the bipartite structure of the ground state, the single band $E_\mathbf{k}=\sqrt{d_\mathbf{k}^2-|\Delta_\mathbf{k}|^2}$ can be viewed as splitting into two bands $E_{1,\mathbf{k}} = E_\mathbf{k}$ and $E_{2,\mathbf{k}}=E_{\mathbf{k}+\mathbf{b}_2/2}$ by the band-folding from the lattice Brillouin zone (BZ) into the magnetic Brillouin zone (${\rm BZ}'$) (See Fig.~\ref{fig:BZ-chi_min}(a).) This can be verified by introducing explicit sublattice labels for the HP bosons, \textit{e.g.} $A$ for even $i_2$ and $B$ for odd $i_2$. Labeling the magnetic unit cell containing site $i$ by $i'$, the Fourier transforms on each sublattice read
\begin{align*}
    a_{i'}^{(A,B)} = \sqrt{\frac{2}{N}} \sum_{\mathbf{k}\in{\rm BZ}'} e^{i\mathbf{k}\cdot\mathbf{R}_{i'}}a_\mathbf{k}^{(A,B)}, \, \mathbf{R}_{i'} = \begin{cases} \mathbf{R}_i, & i\in A, \\ \mathbf{R}_i-\mathbf{a}_2, & i\in B. \end{cases}
\end{align*}
The inverse Fourier transform on the original triangular lattice gives the relations between $a_\mathbf{k}$, $a_{\mathbf{k}+\mathbf{b}_2/2}$, and $a_\mathbf{k}^{(A,B)}$:
\begin{align*}
    a_\mathbf{k} =& \frac{1}{\sqrt{N}} \sum_{i'} e^{-i\mathbf{k}\cdot\mathbf{R}_{i'}}(a_{i'}^{(A)}+e^{-i\mathbf{k}\cdot\mathbf{a}_2}a_{i'}^{(B)})
    = \frac{a_\mathbf{k}^{(A)}+e^{-i\mathbf{k}\cdot\mathbf{a}_2}a_\mathbf{k}^{(B)}}{\sqrt{2}}, \\
    a_{\mathbf{k}+\mathbf{b}_2/2} =& \frac{1}{\sqrt{N}} \sum_{i'} e^{-i\mathbf{k}\cdot\mathbf{R}_{i'}}(a_{i'}^{(A)}-e^{-i\mathbf{k}\cdot\mathbf{a}_2}a_{i'}^{(B)})
    = \frac{a_\mathbf{k}^{(A)}-e^{-i\mathbf{k}\cdot\mathbf{a}_2}a_\mathbf{k}^{(B)}}{\sqrt{2}}.
\end{align*}
The $4\times 4$ magnon Hamiltonian in the sublattice degrees of freedom $a_\mathbf{k}^{(A,B)}$ factorizes into two $2\times 2$ block Hamiltonians of the form $H_2$ in Eq. \eqref{eq:H_full_UCS} when written in terms of $a_\mathbf{k}$ and $a_{\mathbf{k}+\mathbf{b}_2/2}$, which can subsequently be identified with the operators $a_{\sigma,\mathbf{k}}$ for the two magnetic bands $\sigma=1,2$. Instead of summing over $\mathbf{k}\in{\rm BZ}'$ and the magnetic band indices $\sigma=1,2$, we simply sum over the whole BZ in the rest of the paper and keep in mind that
\begin{align*}
    E_\mathbf{k} = \begin{cases} E_{1,\mathbf{k}}, & \mathbf{k}\in{\rm BZ}', \\ E_{2,\mathbf{k}}, & \mathbf{k}\in{\rm BZ}\setminus{\rm BZ}'.\end{cases}
\end{align*}

\begin{figure}[]
    \centering
    \includegraphics[width=0.48\textwidth]{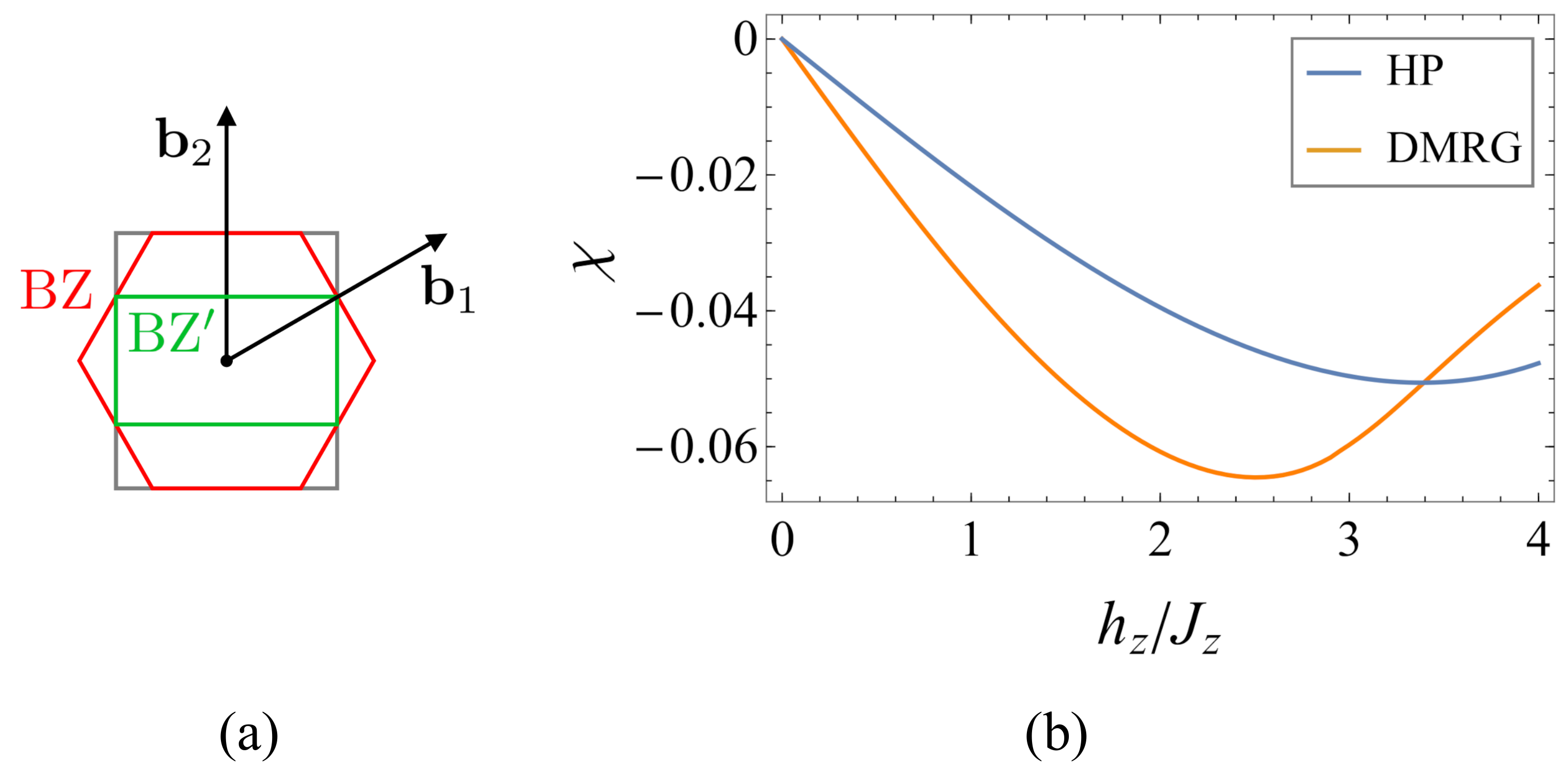}
    \caption{(a) The convention for the lattice Brillouin zone (BZ) and the magnetic Brillouin zone (${\rm BZ}'$) in the UCS phase. (b) QSSC $\chi(J_{\rm PD}=0.5)$ as function of the magnetic field $h_z$, obtained from DMRG and from HP theory \eqref{eq:chi-UCS} at $\alpha=\alpha_0$.}
\label{fig:BZ-chi_min}
\end{figure}

The linear term $H_1$ of the Hamiltonian vanishes at the classical canting angle $\alpha=\alpha_0$ defined in Eq. \eqref{eq:alpha0}. For $h_z=0$, $\alpha_0$ is identically zero, and for $h_z\neq 0$, the same equation reproduces about 90\% of the canting angles obtained from DMRG (See Table \ref{tab:alpha}).

Substituting the classical canting angle $\alpha=\alpha_0$ into the free magnon Hamiltonian \eqref{eq:H_full_UCS}, one obtains Eq. \eqref{eq:deltaS-UCS} for the spin fluctuation $\delta S$ and Eq. \eqref{eq:chi-UCS} for the QSSC $\chi_2$ computed to quadratic order in the magnon expansion. These expressions approximate the DMRG results with relative errors of order 40\% (See Tables \ref{tab:deltaS-UCS} and \ref{tab:chi-UCS}). For $J_{\rm PD}=0.5J_z$, the minimum in QSSC $\chi$ with respect to $h_z$ shifts from $h_z=2.50J_z$ in the DMRG calculation to $h_z=3.38J_z$ in the HP analysis, \textit{i.e.} beyond the phase boundary with the polarized phase (See Fig. \ref{fig:BZ-chi_min}(b)).

\begin{table}[t]
\begin{tabular}{@{\hspace{0.5cm}}c@{\hspace{0.5cm}} || @{\hspace{0.5cm}}c@{\hspace{0.5cm}} | @{\hspace{0.5cm}}c@{\hspace{0.5cm}} | @{\hspace{0.5cm}}c@{\hspace{0.5cm}} | @{\hspace{0.5cm}}c@{\hspace{0.5cm}}}
\hline
$(h_z,J_{\rm PD})$ (in $J_z$) & $\alpha_0$ ($^\circ$) & $\alpha_2$ ($^\circ$) & $\alpha_3$ ($^\circ$) & $\alpha_{\rm DMRG}$ ($^\circ$) \\ \hline
$(0.5,0.5)$ & $5.12$ & $5.21$ & $5.35$ & $5.52$ \\
$(1.0,0.5)$ & $10.29$ & $10.48$ & $10.76$ & $11.23$ \\
$(2.0,0.4)$ & $22.62$ & $23.11$ & $23.87$ & $25.11$ \\
$(2.5,0.3)$ & $31.39$ & $32.20$ & $33.34$ & $34.74$ \\ \hline
\end{tabular}
\caption{The canting angles obtained from the classical configuration ($\alpha_0$), including the quantum corrections of the HP analysis ($\alpha_{2,3}$), and from DMRG ($\alpha_{\rm DMRG}$).}
\label{tab:alpha}
\end{table}
\begin{table}[t]
\begin{tabular}{@{\hspace{0.5cm}}c@{\hspace{0.5cm}} || @{\hspace{0.5cm}}c@{\hspace{0.5cm}} | @{\hspace{0.5cm}}c@{\hspace{0.5cm}} | @{\hspace{0.5cm}}c@{\hspace{0.5cm}} | @{\hspace{0.5cm}}c@{\hspace{0.5cm}}}
\hline
$(h_z,J_{\rm PD})$ (in $J_z$) & $\delta S(\alpha_0)$ & $\delta S(\alpha_2)$ & $\delta S(\alpha_3)$ & $\delta S_{\rm DMRG}$ \\ \hline
$(0.5,0.5)$ & $0.0143$ & $0.0144$ & $0.0144$ & $0.0228$ \\
$(1.0,0.5)$ & $0.0152$ & $0.0153$ & $0.0155$ & $0.0257$ \\
$(2.0,0.4)$ & $0.0239$ & $0.0248$ & $0.0262$ & $0.0443$ \\
$(2.5,0.3)$ & $0.0340$ & $0.0371$ & $0.0430$ & $0.0568$ \\ \hline
\end{tabular}
\caption{Zero-temperature spin fluctuation $\delta S$ computed at the classical canting angles $\alpha=\alpha_0$ using Eq. \eqref{eq:deltaS-UCS}, at the quantum-corrected values $\alpha=\alpha_{2,3}$ using Eq. \eqref{eq:deltaS2-UCS}, and the corresponding DMRG data $\delta S_{\rm DMRG}$.}
\label{tab:deltaS-UCS}
\end{table}
\begin{table}[h]
\begin{tabular}{@{\hspace{0.5cm}}c@{\hspace{0.5cm}} || @{\hspace{0.5cm}}c@{\hspace{0.5cm}} | @{\hspace{0.5cm}}c@{\hspace{0.5cm}} | @{\hspace{0.5cm}}c@{\hspace{0.5cm}} | @{\hspace{0.5cm}}c@{\hspace{0.5cm}}}
\hline
$(h_z,J_{\rm PD})$ (in $J_z$) & $\chi_2(\alpha_0$) & $\chi(\alpha_2$) & $\chi(\alpha_3)$ & $\chi_{\rm DMRG}$ \\ \hline
$(0.5,0.5)$ & $-0.0111$ & $-0.0110$ & $-0.0113$ & $-0.0190$ \\
$(1.0,0.5)$ & $-0.0217$ & $-0.0216$ & $-0.0221$ & $-0.0365$ \\
$(2.0,0.4)$ & $-0.0415$ & $-0.0426$ & $-0.0423$ & $-0.0585$ \\
$(2.5,0.3)$ & $-0.0485$ & $-0.0559$ & $-0.0505$ & $-0.0548$ \\ \hline
\end{tabular}
\caption{Zero-temperature QSSC $\chi$ computed in the free magnon theory to the second-order precision ($\chi_2$) at $\alpha=\alpha_0$, to all orders ($\chi = \chi_2+\chi'_2+\chi_4$) at $\alpha=\alpha_{2,3}$, and from DMRG ($\chi_{\rm DMRG}$).}
\label{tab:chi-UCS}
\end{table}
Including the quantum corrections $E_2=2E_0+\frac{1}{2}Nh_z\sin\alpha + \frac{1}{2}\sum_{\mathbf{k}\in{\rm BZ}} E_\mathbf{k}$ in the minimization of the ground-state energy shifts the optimal angle away from $\alpha_0$. At this quantum-corrected angle, the linear term $H_1$ no longer vanishes. Although this leads to a finite ground-state expectation value $\langle a_\mathbf{k}\rangle$ only at a single crystal momentum $\mathbf{k}=\mathbf{b}_2/2$, it scales as the linear system size $\sqrt{N}$ and generates a transversal spin fluctuation,
\begin{align*}
    \langle \tilde{S}_i^y\rangle = (-1)^{i_2}\frac{(J+3J_z+4J_{\rm PD})(\sin\alpha-\sin\alpha_0)}{2(d_{\mathbf{b}_2/2}-\Delta_{\mathbf{b}_2/2})}\cos\alpha,
\end{align*}
as well as an additional correction to the ground-state energy,
\begin{align*}
    E_{2}' = N \frac{[(J+3J_z+4J_{\rm PD})(\sin\alpha-\sin\alpha_0)]^2}{2(d_{\mathbf{b}_2/2}-\Delta_{\mathbf{b}_2/2})}\cos^2 \alpha.
\end{align*}
Introducing the uniform parameter $v=(-1)^{i_2}\langle \tilde{S}_i^y\rangle$, the expression for the spin fluctuation is modified to
\begin{align}
    \delta S = \frac{1}{2}-\bigg[v^2+\Big\{\frac{1}{2}-\frac{1}{N}\sum_{\mathbf{k}\in{\rm BZ}}  \left( \sinh^2\theta_\mathbf{k} + n_\mathbf{k}(T)\cosh{2\theta_\mathbf{k}} \right)-v^2\Big\}^2\bigg]^{1/2},
\label{eq:deltaS2-UCS}
\end{align}
and the formula \eqref{eq:chi-UCS} for $\chi_2$ acquires additional terms
\begin{align}
    \chi'_2 = \frac{1}{2N} \bigg[\Big(\frac{3}{2}v-v^3\Big)\sin{2\alpha}+v^2(1-3\cos{2\alpha})\bigg]  \sum_{\mathbf{k}\in{\rm BZ}} (2n_\mathbf{k}(T)+1)\sin\psi_\mathbf{k} \alpha \cos(\mathbf{k}\cdot\mathbf{a}_2)\sinh{2\theta_\mathbf{k}}.
\end{align}
Of a comparable magnitude as $\chi'_2$ is the fourth-order contribution $\chi_4$ to the QSSC. At $T=0$,
\begin{align}
    \chi_4 =& \frac{1}{N^2}\sum_{\mathbf{k},\mathbf{q}\in{\rm BZ}} \sin{\psi_\mathbf{k}}\sinh{2\theta_\mathbf{k}}\sinh^2\theta_\mathbf{q}[\cos{(\mathbf{k}\cdot\mathbf{a}_1)}\cos{(\mathbf{q}\cdot\mathbf{a}_2)}+\cos{(\mathbf{q}\cdot\mathbf{a}_1)}\cos{(\mathbf{k}\cdot\mathbf{a}_2)}+\cos{(\mathbf{k}\cdot\mathbf{a}_3)}](\cos^2\alpha+v\sin{2\alpha}) \nn
    & \qquad \quad -\frac{1}{2}\sin{(\psi_\mathbf{k}-\psi_\mathbf{q})}\sinh{2\theta_\mathbf{k}}\sinh{2\theta_\mathbf{q}}\cos{(\mathbf{k}\cdot\mathbf{a}_1)}\cos{(\mathbf{q}\cdot\mathbf{a}_2)}(\sin^2\alpha-v\sin{2\alpha}).
\end{align}
Minimizing the total ground-state energy $E_0+E_2+E_2'$ of the free magnons yields a marginally improved agreement with DMRG for $\alpha$ and $\delta S$, while quantum corrections to $\chi_2(\alpha_0)$ largely cancel out because $\chi_4$ enters with the opposite sign as $\chi_2$ (See Tables \labelcref{tab:alpha,tab:deltaS-UCS,tab:chi-UCS}).

Alternatively, at zero temperature, one can determine the quantum correction to the canting angle by including the interaction terms via a mean-field treatment and requiring that the linear terms in the effective Hamiltonian cancel~\cite{bader25,kesharpu25}. The mean-field treatment of the three-magnon interaction $H_3$ is achieved by the Hartree-Fock decoupling
\begin{align*}
    a_ja_i^\dagger a_i \rightarrow m_\mathbf{\delta}a_i+\Delta_\mathbf{\delta}a^\dagger_i+ \rho a_j,\quad \mathbf{\delta}=\mathbf{R}_j-\mathbf{R}_i,
\end{align*}
where the Hartree-Fock averages are given by
\begin{align}
    m_\mathbf{\delta} &= \frac{1}{N}\sum_i \langle a_i^\dagger a_j\rangle = \frac{1}{N} \sum_{\mathbf{k}\in{\rm BZ}} \sinh^2{\theta_\mathbf{k}}\cos(\mathbf{k}\cdot\mathbf{\delta}), \nn
    \Delta_\mathbf{\delta} &= \frac{1}{N}\sum_i \langle a_i a_j\rangle = \frac{-1}{2N} \sum_{\mathbf{k}\in{\rm BZ}} e^{i\psi_\mathbf{k}}\sinh{2\theta_\mathbf{k}}\cos(\mathbf{k}\cdot\mathbf{\delta}), \nn
    \rho &= \frac{1}{N}\sum_i \langle a_i^\dagger a_i\rangle = \frac{1}{N} \sum_{\mathbf{k}\in{\rm BZ}} \sinh^2{\theta_\mathbf{k}}.
\end{align}
This leads to a linear quantum correction to the free-magnon Hamiltonian,
\begin{align}
    \overline{H}_3 = i\sqrt{N} \bigg\lbrace \frac{J_z-J}{2}\sin{2\alpha} \Big[2\rho-\sum_{n=1}^3 \big( m_{\mathbf{a}_n}-{\rm Re}\Delta_{\mathbf{a}_n} \big) \Big]  -J_{\rm PD}\sin{2\alpha} \Big[ 2\rho+\sum_{n=1}^3 C_n \big( m_{\mathbf{a}_n}-{\rm Re}\Delta_{\mathbf{a}_n} \big) \Big] \bigg\rbrace (a_{\mathbf{b}_2/2}-a_{\mathbf{b}_2/2}^\dagger),
\end{align}
and a condition for the renormalized canting angle $\alpha_3$,
\begin{align}
    H_1(\alpha=\alpha_3)+\overline{H}_3(\alpha=\alpha_3)=0.
\end{align}
While the condition $H_1(\alpha)=0$ is equivalent to the classical configuration $\alpha=\alpha_0$, $\alpha_3$ differs in general from the angle $\alpha_2$ that minimizes the total ground-state energy of the free-magnon Hamiltonian. The spin fluctuation $\delta S$ and the QSSC $\chi$ can be computed using the same formulae as with $\alpha_2$, except that now $v=0$ by assumption. The quantum correction for the canting angle and $\delta S$ provided by $\alpha_3$ is twice the amount of the correction from $\alpha_2$, while the result for QSSC is not necessarily improved by using $\alpha_3$ instead of $\alpha_2$.

\section{HP analysis of the polarized phases}
\label{app:P-phase}

Since the spin expectation values at every lattice site already point in the z-direction, it is straightforward to derive the HP Hamiltonian for the polarized (P) phases. For the P phase in the phase diagram of $H_{\rm XXZ} + H_{\rm PD}$, we get
\begin{align}
    E_0 =& N\bigg(\frac{3}{4}J_z-\frac{h_z}{2}\bigg), \quad H_1 = 0, \nn
    H_2 =& \sum_{\mathbf{k}\in{\rm BZ}} E_\mathbf{k} \left(b_\mathbf{k}^\dagger b_\mathbf{k} + \frac{1}{2}\right) +\frac{3J_z-h_z}{2}N, \quad E_\mathbf{k} = \sqrt{d_\mathbf{k}^2-|\Delta_\mathbf{k}|^2}, \nn
    d_\mathbf{k} =& J\sum_{n=1}^3 \cos{(\mathbf{k}\cdot\mathbf{a}_n)}-3J_z+h_z, \nn
    \Delta_\mathbf{k} =& 2J_{\rm PD} \bigg[ \sum_{n=1}^3 C_n\cos{(\mathbf{k}\cdot\mathbf{a}_n)} -i\frac{\sqrt{3}}{2}[\cos{(\mathbf{k}\cdot\mathbf{a}_2)}-\cos{(\mathbf{k}\cdot\mathbf{a}_3)}] \bigg],
\end{align}
with a Bogoliubov transformation of the form \eqref{eq:Bogoliubov_UCS} that relates the HP bosons $a_\mathbf{k}, a_\mathbf{k}^\dagger$ and the quasiparticle excitations $b_\mathbf{k}, b_\mathbf{k}^\dagger$.

By equating the classical ground-state energies of the UCS and the P phases, $E_{0,{\rm UCS}}(\alpha=\alpha_0)=E_{0,{\rm P}}$, we find the mean-field estimate of the UCS-P phase boundary,
\begin{align*}
    J_{\rm PD} = \frac{1}{4}(h_z-J-3J_z),
\end{align*}
which is precisely where the classical canting angle $\alpha_0$ of the UCS phase becomes $90^\circ$, but it does not match the phase boundary obtained from DMRG very well. It is not possible to improve the identification of the phase boundary by including quantum corrections from $H_{1, \rm UCS}$ and $H_{2, \rm UCS/P}$, as the regions in the phase diagram where the HP theories of the two phases are valid, given by the real-valuedness of the excitation energy $E_\mathbf{k}=\sqrt{d_\mathbf{k}^2-|\Delta_\mathbf{k}|^2}$ for all $\mathbf{k}\in{\rm BZ}$, do not intersect (\textit{cf:} Fig. \ref{fig:UCS-P_bound}).
\begin{figure}[]
    \centering
    \includegraphics[width=0.45\textwidth]{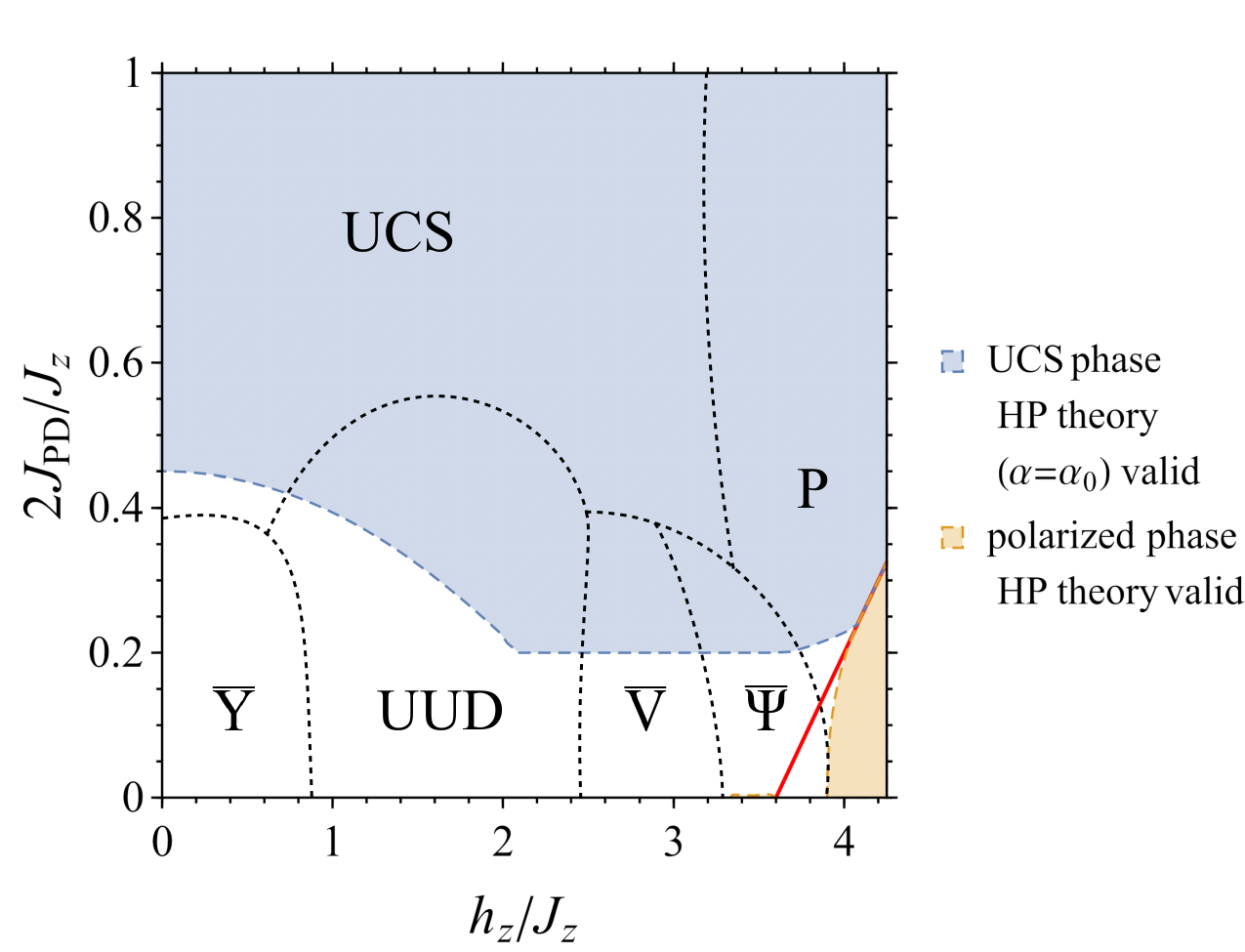}
    \caption{Regions of validity for the HP theories of the UCS the P phases ($J_\Gamma=0$). The red straight line is the mean-field estimate of the UCS-P phase boundary, and the black dashed lines indicate the phase boundaries determined by DMRG.}
    \label{fig:UCS-P_bound}
\end{figure}

The spin fluctuation $\delta S$ is given by the same formula as in the UCS phase, Eq. \eqref{eq:deltaS-UCS}, yielding $\delta S=0.0079$ for $(h_z,J_{\rm PD})=(4.25,0.05)J_z$. Although this underestimates the DMRG result $\delta S=0.0219$ by more than a factor of two, both values are sufficiently small compared to $S=1/2$ to justify the use of free-magnon theory.

The QSSC can be separated into terms of second and fourth orders in the magnon expansion,
\begin{align*}
    \chi = \chi(S_i^z=1/2)+\chi(S_i^z=-a_i^\dagger a_i) \equiv \chi_2+\chi_4.
\end{align*}
The second-order contribution $\chi_2$ to the QSSC turns out to vanish strictly,
\begin{align*}
    \chi_2 =& \frac{-i}{8N} \sum_i \langle 
    a_{\mathbf{R}_i}^\dagger a_{\mathbf{R}_i+\mathbf{a}_2}
    -a_{\mathbf{R}_i+\mathbf{a}_2}^\dagger a_{\mathbf{R}_i}
    +a_{\mathbf{R}_i+\mathbf{a}_2}^\dagger a_{\mathbf{R}_i-\mathbf{a}_1} -a_{\mathbf{R}_i-\mathbf{a}_1}^\dagger a_{\mathbf{R}_i+\mathbf{a}_2}
    +a_{\mathbf{R}_i-\mathbf{a}_1}^\dagger a_{\mathbf{R}_i}
    -a_{\mathbf{R}_i}^\dagger a_{\mathbf{R}_i-\mathbf{a}_1} +(\mathbf{a}_1\rightarrow-\mathbf{a}_1,\,\mathbf{a}_2\rightarrow-\mathbf{a}_2)\rangle\nn
    =& \frac{1}{4N}\sum_{\mathbf{k}\in{\rm BZ}} \langle a_{\mathbf{k}}^\dagger a_\mathbf{k}\rangle \sum_{n=1}^3 \sin{(\mathbf{k}\cdot\mathbf{a}_n)}+\sin{(-\mathbf{k}\cdot\mathbf{a}_n)}=0,
\end{align*}
which is consistent with the no-go theorem~\cite{katsura10} that is valid up to the second order in $\delta \mathbf{S}_i=\mathbf{S}_i-\langle\mathbf{S}_i\rangle_{\rm cl}$.

A nonzero QSSC arises from the fourth-order contribution $\chi_4$, which coincides with the third-order contribution in $\delta \Sp_i$,
\begin{align}
    \chi_4 =& \frac{1}{2N}\sum_i \langle \delta\Sp_{\mathbf{R}_i} \cdot \delta\Sp_{\mathbf{R}_i+\mathbf{a}_2} \times \delta\Sp_{\mathbf{R}_i-\mathbf{a}_1} + \delta\Sp_{\mathbf{R}_i} \cdot \delta\Sp_{\mathbf{R}_i-\mathbf{a}_2} \times \delta\Sp_{\mathbf{R}_i+\mathbf{a}_1}\rangle \nn
    =& \frac{i}{4N} \sum_i \langle 
    (a_{\mathbf{R}_i}^\dagger a_{\mathbf{R}_i+\mathbf{a}_2}
    -a_{\mathbf{R}_i+\mathbf{a}_2}^\dagger a_{\mathbf{R}_i})
    a_{\mathbf{R}_i-\mathbf{a}_1}^\dagger a_{\mathbf{R}_i-\mathbf{a}_1} 
    +(a_{\mathbf{R}_i+\mathbf{a}_2}^\dagger a_{\mathbf{R}_i-\mathbf{a}_1}
    -a_{\mathbf{R}_i-\mathbf{a}_1}^\dagger a_{\mathbf{R}_i+\mathbf{a}_2})
    a_{\mathbf{R}_i}^\dagger a_{\mathbf{R}_i} \nn
    &\qquad +(a_{\mathbf{R}_i-\mathbf{a}_1}^\dagger a_{\mathbf{R}_i}
    -a_{\mathbf{R}_i}^\dagger a_{\mathbf{R}_i-\mathbf{a}_1})
    a_{\mathbf{R}_i+\mathbf{a}_2}^\dagger a_{\mathbf{R}_i+\mathbf{a}_2}
    +(\mathbf{a}_1\rightarrow-\mathbf{a}_1,\,\mathbf{a}_2\rightarrow-\mathbf{a}_2)\rangle.
\label{eq:chi4_P_1}
\end{align}
Performing the Fourier and the Bogoliubov transformations, each 4-point function of the HP bosons can be written as a linear combination of magnon 4-point functions. At zero temperature, we obtain
\begin{align}
    \langle a_{\mathbf{R}_i+\mathbf{r}_1}^\dagger a_{\mathbf{R}_i+\mathbf{r}_2}
    a_{\mathbf{R}_i+\mathbf{r}_3}^\dagger a_{\mathbf{R}_i+\mathbf{r}_3}\rangle =& \frac{1}{N}\sum_{\mathbf{k},\mathbf{p},\mathbf{q}} e^{i[\mathbf{k}\cdot(\mathbf{r}_3-\mathbf{r}_1)+\mathbf{p}\cdot(\mathbf{r}_2-\mathbf{r}_3)]} \langle a_\mathbf{k}^\dagger a_\mathbf{p}
    a_\mathbf{q}^\dagger a_{\mathbf{k}-\mathbf{p}+\mathbf{q}}\rangle \nn
    =& \frac{1}{N} \sum_{\mathbf{k},\mathbf{p}} \bigg\lbrace e^{i[\mathbf{k}\cdot(\mathbf{r}_3-\mathbf{r}_1)+\mathbf{p}\cdot(\mathbf{r}_2-\mathbf{r}_3)]} \Big[\frac{1}{4}e^{i(\psi_\mathbf{p}-\psi_\mathbf{k})}\sinh{2\theta_\mathbf{k}}\sinh{2\theta_\mathbf{p}}+\sinh^2\theta_\mathbf{k}\cosh^2\theta_\mathbf{p}\Big] \nn
    & \mkern45mu +e^{i\mathbf{k}\cdot(\mathbf{r}_2-\mathbf{r}_1)}\sinh^2\theta_\mathbf{k}\sinh^2\theta_\mathbf{p} \bigg\rbrace,
\label{eq:chi4_P_2}
\end{align}
where we used the fact that the only non-vanishing magnon 4-point functions are
\begin{align*}
    \langle b_{-\mathbf{k}} b_\mathbf{p}
    b_\mathbf{q}^\dagger b_{-\mathbf{k}+\mathbf{p}-\mathbf{q}}^\dagger\rangle = \delta_{\mathbf{k},-\mathbf{q}} +\delta_{\mathbf{p},\mathbf{q}}, \quad
    \langle b_{-\mathbf{k}} b_{-\mathbf{p}}^\dagger
    b_{-\mathbf{q}} b_{-\mathbf{k}+\mathbf{p}-\mathbf{q}}^\dagger\rangle = \delta_{\mathbf{k},\mathbf{p}}.
\end{align*}
Substituting Eq. \eqref{eq:chi4_P_2} into Eq. \eqref{eq:chi4_P_1} yields
\begin{align}
    \chi_4=\frac{-i}{2N^2} \sum_{\mathbf{k},\mathbf{p}} \begin{vmatrix}
        \cos{(\mathbf{k}\cdot\mathbf{a}_1)} & \cos{(\mathbf{k}\cdot\mathbf{a}_2)} & \cos{(\mathbf{k}\cdot\mathbf{a}_3)} \\
        \cos{(\mathbf{p}\cdot\mathbf{a}_1)} & \cos{(\mathbf{p}\cdot\mathbf{a}_2)} & \cos{(\mathbf{p}\cdot\mathbf{a}_3)} \\
        1 & 1 & 1
    \end{vmatrix} \Big[\frac{1}{4}e^{i(\psi_\mathbf{p}-\psi_\mathbf{k})}\sinh{2\theta_\mathbf{k}}\sinh{2\theta_\mathbf{p}}+\sinh^2\theta_\mathbf{k}\cosh^2\theta_\mathbf{p}\Big].
\label{eq:chi4_P_3}
\end{align}
Here, we used that the Bogoliubov angles $\psi_\mathbf{k},\,\theta_\mathbf{k}$ are symmetric under $\mathbf{k}\rightarrow-\mathbf{k}$, so that we can replace $e^{i[\mathbf{k}\cdot(\mathbf{r}_3-\mathbf{r}_1)+\mathbf{p}\cdot(\mathbf{r}_2-\mathbf{r}_3)]}\rightarrow \cos(\mathbf{k}\cdot(\mathbf{r}_3-\mathbf{r}_1))\cos(\mathbf{p}\cdot(\mathbf{r}_2-\mathbf{r}_3))$ and the contributions from the third line of Eq. \eqref{eq:chi4_P_2} cancel out. Eq. \eqref{eq:chi4_P_3} can be further reduced by exploiting the symmetry properties under $k_y\rightarrow -k_y$, which induces the change $\mathbf{k}\cdot\mathbf{a}_2\leftrightarrow \mathbf{k}\cdot\mathbf{a}_3$, $\Delta_\mathbf{k}\rightarrow \Delta_\mathbf{k}^*$, and $\psi_\mathbf{k}\rightarrow -\psi_\mathbf{k}$. This implies that, under $(k_y,p_y)\rightarrow (-k_y,-p_y)$, the determinant factor and the second term in the squared bracket are antisymmetric and symmetric, respectively, while the first term in the squared bracket is mapped onto its complex conjugate. Finally, we get
\begin{align}
    \chi_4 = -\frac{1}{8N^2} \sum_{\mathbf{k},\mathbf{p}} \begin{vmatrix}
        \cos{(\mathbf{k}\cdot\mathbf{a}_1)} & \cos{(\mathbf{k}\cdot\mathbf{a}_2)} & \cos{(\mathbf{k}\cdot\mathbf{a}_3)} \\
        \cos{(\mathbf{p}\cdot\mathbf{a}_1)} & \cos{(\mathbf{p}\cdot\mathbf{a}_2)} & \cos{(\mathbf{p}\cdot\mathbf{a}_3)} \\
        1 & 1 & 1
    \end{vmatrix} \sin{(\psi_\mathbf{k}-\psi_\mathbf{p})}\sinh{2\theta_\mathbf{k}}\sinh{2\theta_\mathbf{p}}.
\label{eq:chi4_P}
\end{align}

With $\chi_4(h_z=4.25J_z,J_{\rm PD}=0.05J_z)=-0.0033$, the HP-theory prediction exceeds the DMRG result $-0.0009$ by more than a factor of three in magnitude. However, the parameter region of the P phase currently accessible to both HP theory and DMRG lies rather close to the boundary beyond which HP theory breaks down, and the discrepancy between the HP and DMRG results tends to decrease as one moves away from this boundary. Eq. \eqref{eq:chi4_P} therefore provides at least a qualitatively reliable account of the nonzero QSSC in the P phase.

For the P phase in the phase diagram of $H_{\rm XXZ} + H_\Gamma$, the free part of the magnon Hamiltonian is already diagonal in terms of the HP bosons $a_\mathbf{k}$ because $H_\Gamma$ contains only three-magnon terms. Since both the QSSC and the Berry curvature can be nonzero only for ${\rm Im}\Delta_\mathbf{k}\neq 0$, they vanish in this P phase.

\section{HP analysis of the canted stripe phase}
\label{app:CS-phase}

Using the spin parameterization \eqref{eq:CS}, we evaluate the classical (zeroth-order) ground-state energy of the canted stripe (CS) configuration,
\begin{align}
    \frac{E_0}{N}
    =& \frac{J}{8}\!\left(\sin^2\Theta_A + \sin^2\Theta_B - 4\sin\Theta_A\sin\Theta_B\right)
      + \frac{J_z}{8}\!\left(\cos^2\Theta_A + \cos^2\Theta_B - 4\cos\Theta_A\cos\Theta_B\right) \nn
    & - \frac{J_\Gamma}{4}\sin\!\Big(\phi+\frac{\pi}{3}\Big)
        \Bigl[\sin(\Theta_A+\Theta_B)+\frac{1}{2}(\sin2\Theta_A+\sin2\Theta_B)\Bigr] - \frac{h_z}{4}\!\left(\cos\Theta_A - \cos\Theta_B\right)
        ,
    \label{eq:E0_general_phi}
\end{align}
where we have denoted by $\Theta_A$ and $\Theta_B$ the polar angles of the two sublattices with respect to the global $z$ axis. Among all terms in Eq.~\eqref{eq:E0_general_phi}, only the $J_\Gamma$ term depends on $\phi$. Minimizing the energy with respect to $\phi$ gives $\phi=\pi/6$.

As in the UCS phase, there are three degenerate ground states in the CS phase. The other two are obtained by setting $\eta_i$ in the classical spin configuration \eqref{eq:CS} to $(-1)^{i_2}$ or $(-1)^{i_1+i_2}$, and repeating the energy minimization, which yields $\phi=-\pi/2$ or $5\pi/6$, respectively. Unlike the ground states of the UCS phase, those of the CS phase are not invariant under a $\pi$ rotation about the $z$-axis, which reflects the symmetry structures of the corresponding Hamiltonians $H_{\rm XXZ}+H_{\rm PD}$ and $H_{\rm XXZ}+H_\Gamma$.

In the absence of the magnetic field ($h_z=0$), the canted stripe configuration is characterized by a single polar angle $\Theta_A=\Theta_B=\theta$ that specifies the common canting axis of the two sublattices. For $\phi=\pi/6$, the classical energy per site \eqref{eq:E0_general_phi} becomes
\begin{align*}
    \frac{E_0}{N}
    = -\frac{1}{4}\!\left(
        J \sin^2\theta
        + J_z \cos^2\theta
      \right)
      - \frac{J_\Gamma}{2}\sin 2\theta .
\end{align*}
Minimizing this with respect to $\theta$ yields Eq. \eqref{eq:theta_h=0}. Under a finite out-of-plane field ($h_z\neq0$), the stationary conditions $\partial E_0/\partial\Theta_A=0$ and $\partial E_0/\partial\Theta_B=0$ lead to the coupled equations
\begin{align*}
0 &= \frac{J-J_z}{8}\sin(2\Theta_A)
   + \frac{J+J_z}{4}\sin(\Theta_A-\Theta_B)
   + \frac{J_z-J}{4}\sin(\Theta_A+\Theta_B) \nonumber\\
  &\quad + \frac{h_z}{4}\sin\Theta_A
   - \frac{J_\Gamma}{4}
     \!\left[\cos(2\Theta_A)+\cos(\Theta_A+\Theta_B)\right],
   \nn[3pt]
0 &= \frac{J-J_z}{8}\sin(2\Theta_B)
   - \frac{J+J_z}{4}\sin(\Theta_A-\Theta_B)
   + \frac{J_z-J}{4}\sin(\Theta_A+\Theta_B) \nonumber\\
  &\quad - \frac{h_z}{4}\sin\Theta_B
   - \frac{J_\Gamma}{4}
     \!\left[\cos(2\Theta_B)+\cos(\Theta_A+\Theta_B)\right].
\end{align*}
For any finite field $h_z>0$, the two canting angles are generically inequivalent. For $(h_z,J_\Gamma)=(0.2,1.0)J_z$, we find $\Theta_A \simeq 40.70^\circ$ and $\Theta_B \simeq 43.69^\circ$, confirming a small but finite sublattice asymmetry induced by the external field and the $\Gamma$ interaction.

To describe spin-wave excitations around the classical CS configuration, we introduce a local coordinate frame at each lattice site $i$ such that the local $z'$ axis is aligned with the classical spin direction $\langle \mathbf{S}_i \rangle$. The transformation between the laboratory frame $(x,y,z)$ and the local spin frame $(x',y',z')$ is implemented by a site-dependent rotation matrix,
\begin{align*}
R_i 
= R_z(\phi)\,
R_y\!\left(
\Theta_i
\right)\,
R_x\!\left(\tfrac{\pi}{2}(1-\eta_i)\right),
\end{align*}
\pagebreak
where $R_\mu(\alpha)$ denotes a rotation by an angle $\alpha$ about the $\mu$-axis and $\Theta_i$ is as defined in Eq. \eqref{eq:theta_i}. The spin operators in the laboratory and the local coordinate frames $\Sp_i,\,\tilde \Sp_i$ are then related by $\mathbf{S}_i = R_i\,\tilde{\mathbf{S}}_i$, or writing out each component,
\begin{align*}
S_i^x &= 
 \cos\phi \cos\Theta_i\,\tilde S_i^x
 - \eta_i \sin\phi\,\tilde S_i^y
 + \eta_i \cos\phi \sin\Theta_i\,\tilde S_i^z, \nn
S_i^y &= 
 \sin\phi \cos\Theta_i\,\tilde S_i^x
 + \eta_i \cos\phi\,\tilde S_i^y
 + \eta_i \sin\phi \sin\Theta_i\,\tilde S_i^z, \nn
S_i^z &=
 - \sin\Theta_i\,\tilde S_i^x
 + \eta_i \cos\Theta_i\,\tilde S_i^z.
\end{align*}

Applying the HP transformation to the rotated spin operators $\tilde \Sp_i$ and retaining only quadratic terms in the ladder operators from the Hamiltonian $H_{\rm XXZ}+H_\Gamma$, we obtain
\begin{align*}
    H_2=&\sum_{\langle i,j \rangle}\Bigl(A_{ij}(a_i^\dagger a_j + a_i a_j^\dagger  )+B_{ij}(a_i a_j + a_i^\dagger a_j^\dagger)+C_{ij}(a_ia_j^\dagger - a_i^\dagger a_j)+D_{ij}(a_ia_j - a_i^\dagger a_j^\dagger)+E_{ij}(a_i^\dagger a_i + a_j^\dagger a_j)\Bigl) \nn 
    &+ h_z\sum_i \eta_i\cos\Theta_i a_i^\dagger a_i,
\end{align*}
where the coefficients $A_{ij}$, $B_{ij}$, $C_{ij}$, $D_{ij}$, and $E_{ij}$ are given by
\begin{align*}
    A_{ij} &=\frac{J}{4}(\cos\Theta_i\cos\Theta_j+\eta_i\eta_j)+\frac{J_z}{4}\sin\Theta_i\sin\Theta_j+\frac{J_{\Gamma}}{4}\sin{(\Theta_i+\Theta_j)} \sin(\varphi_{ij}-\phi), \nn 
    B_{ij}&=\frac{J}{4}(\cos\Theta_i\cos\Theta_j-\eta_i\eta_j)+\frac{J_z}{4}\sin\Theta_i\sin\Theta_j+\frac{J_{\Gamma}}{4}\sin{(\Theta_i+\Theta_j)} \sin(\varphi_{ij}-\phi), \nn
    C_{ij}&= - \frac{i}{4}J_{\Gamma}(\eta_i\sin\Theta_j-\eta_j\sin\Theta_i)\cos{(\varphi_{ij}-\phi)}, \nn
    D_{ij}&= \frac{i}{4}J_{\Gamma}(\eta_i\sin\Theta_j+\eta_j\sin\Theta_i)\cos{(\varphi_{ij}-\phi)}, \nn
    E_{ij}&=-\frac{\eta_i\eta_j}{2}[J\sin\Theta_i\sin\Theta_j+J_z\cos\Theta_i\cos\Theta_j-J_\Gamma\sin(\Theta_i+\Theta_j)\sin{(\varphi_{ij}-\phi)}].
\end{align*}
Introducing the index $i'$ for the magnetic unit cell containing the site $i$, we replace the ladder operators $a_i^{(\dag)}$ by $a_{i'}^{(\dag)}$ for $i\in A$ and $b_{i'}^{(\dag)}$ for $i\in B$. For $\eta_i=(-1)^{i_1}$ and $\phi=\pi/6$, applying the Fourier transformations
\begin{align*}
    a_{i'} = \sqrt{\frac{2}{N}}\sum_{\mathbf{k}\in {\rm BZ}'}
    e^{i\mathbf{k}\cdot\mathbf{R}_{i}} a_{\mathbf k}, \quad
       b_{i'} = \sqrt{\frac{2}{N}}\sum_{\mathbf{k}\in {\rm BZ}'}
    e^{i\mathbf{k}\cdot(\mathbf{R}_{i}-\mathbf{a}_1)} b_{\mathbf k},
\end{align*}
to the ladder operators yields the magnon Hamiltonian
\begin{align*}
    H_m =
    \sum_{\mathbf{k}\in {\rm BZ}'}
    \Bigl[
        \frac{1}{2}\,
        \Psi_{\mathbf{k}}^\dagger
        H_{\mathbf{k}}
        \Psi_{\mathbf{k}}
        -
        \frac{1}{2}(\epsilon_A+\epsilon_B)
    \Bigr]
    + E_0, \quad 
    H_{\mathbf{k}} =
    \begin{pmatrix}
        h_{\mathbf{k}} & \Delta_{\mathbf{k}} \\
        \Delta^\dagger_{\mathbf{k}} & h^\top_{-\mathbf{k}}
    \end{pmatrix},
\end{align*}
with the Nambu spinor $\Psi_{\mathbf{k}}
    =
    \begin{pmatrix}
        a_{\mathbf{k}} & 
        b_{\mathbf{k}} & 
        a_{-\mathbf{k}}^\dagger &
        b_{-\mathbf{k}}^\dagger
    \end{pmatrix}^\top$. Here, the $2\times2$ matrices $h_\mathbf{k},\,\Delta_\mathbf{k}$ are given by
\begin{align}
h_{\mathbf k} &=
\begin{pmatrix}
2\cos(\mathbf{k}\cdot\mathbf{a}_2)A_2^{(A)}+\epsilon_A
&
A_{1,3} f_{\mathbf k} + C_0 g_{\mathbf k}
\\
A_{1,3} f^{*}_{\mathbf k} + C_0^{*} g^{*}_{\mathbf k}
&
2\cos(\mathbf{k}\cdot\mathbf{a}_2)A_2^{(B)}+\epsilon_B
\end{pmatrix},
\nn
[6pt]
\Delta_{\mathbf k} &=
\begin{pmatrix}
2\cos(\mathbf{k}\cdot\mathbf{a}_2)B_2^{(A)}
&
B_{1,3} f_{\mathbf k} + D_0^{*} g_{\mathbf k}
\\
B_{1,3} f_{-\mathbf k} + D_0^{*} g_{-\mathbf k}
&
2\cos(\mathbf{k}\cdot\mathbf{a}_2)B_2^{(B)}
\end{pmatrix},
\end{align}
and the appearing coefficients are defined as
\begin{align}
    A_2^{(A)} &\equiv A_{ij}\big\rvert_{i,j\in A}= \frac{J}{4}(\cos^2\Theta_A+1)+\frac{J_z}{4}\sin^2\Theta_A+\frac{J_\Gamma}{4}\sin{2\Theta_A}, \nn
    A_2^{(B)} &\equiv A_{ij}\big\rvert_{i,j\in B}= \frac{J}{4}(\cos^2\Theta_B+1)+\frac{J_z}{4}\sin^2\Theta_B+\frac{J_\Gamma}{4}\sin{2\Theta_B}, \nn
    A_{1,3} &\equiv A_{ij}\big\rvert_{i\in A,j\in B}= A_{ij}\big\rvert_{i\in B,j\in A} \nn
    &= \frac{J}{4}(\cos\Theta_A\cos\Theta_B-1)+\frac{J_z}{4}\sin\Theta_A\sin\Theta_B-\frac{J_\Gamma}{8}\sin{(\Theta_A+\Theta_B)}, \nn
    B_2^{(A)}&\equiv B_{ij}\big\rvert_{i,j\in A}=\frac{J}{4}(\cos^2\Theta_A-1)+\frac{J_z}{4}\sin^2\Theta_A+\frac{J_\Gamma}{4}\sin{2\Theta_A}, \nn
    B_2^{(B)}&\equiv B_{ij}\big\rvert_{i,j\in B}=\frac{J}{4}(\cos^2\Theta_B-1)+\frac{J_z}{4}\sin^2\Theta_B+\frac{J_\Gamma}{4}\sin{2\Theta_B}, \nn
    B_{1,3}&\equiv B_{ij}\big\rvert_{i\in A,j\in B}= B_{ij}\big\rvert_{i\in B,j\in A} \nn
    &=\frac{J}{4}(\cos\Theta_A\cos\Theta_B+1)+\frac{J_z}{4}\sin\Theta_A\sin\Theta_B-\frac{J_\Gamma}{8}\sin(\Theta_A+\Theta_B), \nn
    C_0&\equiv C_{i,j}\big\rvert_{i\in A,\varphi_{ij}=0} = -C_{i,j}\big\rvert_{i\in B,\varphi_{ij}=0} = -C_{i,j}\big\rvert_{i\in A,\varphi_{ij}=-2\pi/3} = C_{i,j}\big\rvert_{i\in B,\varphi_{ij}=-2\pi/3} \nn
    &=-i\frac{J_\Gamma\sqrt{3}}{8}(\sin\Theta_A+\sin\Theta_B), \nn
    D_0&\equiv D_{i,j}\big\rvert_{i\in A,\varphi_{ij}=0} = -D_{i,j}\big\rvert_{i\in B,\varphi_{ij}=0} = -D_{i,j}\big\rvert_{i\in A,\varphi_{ij}=-2\pi/3} = D_{i,j}\big\rvert_{i\in B,\varphi_{ij}=-2\pi/3} \nn
    &=i\frac{J_\Gamma\sqrt{3}}{8}(\sin\Theta_B - \sin\Theta_A), \nn
    E^{(A)}&\equiv 2\Big( E_{ij}\big\rvert_{i,j\in A} + E_{ij}\big\rvert_{i\in A,j\in B} + E_{ij}\big\rvert_{i\in B,j\in A} \Big) \nn
    &=J(2\sin{\Theta_A}\sin{\Theta_B}-\sin^2{\Theta_A})+J_z(2\cos{\Theta_A}\cos{\Theta_B}-\cos^2{\Theta_A})+J_\Gamma\bigl(\sin(\Theta_A+\Theta_B)+\sin(2\Theta_A)\bigr), \nn
    E^{(B)}&\equiv 2\Big( E_{ij}\big\rvert_{i,j\in B} + E_{ij}\big\rvert_{i\in A,j\in B} + E_{ij}\big\rvert_{i\in B,j\in A} \Big) \nn
    &=J(2\sin{\Theta_A}\sin{\Theta_B}-\sin^2{\Theta_B})+J_z(2\cos{\Theta_A}\cos{\Theta_B}-\cos^2{\Theta_B})+J_\Gamma\bigl(\sin(\Theta_A+\Theta_B)+\sin(2\Theta_B)\bigr), \nn
    \epsilon_A&=E^{(A)} + h_z\cos\Theta_A,\quad \epsilon_B=E^{(B)} - h_z\cos\Theta_B, \nn
    f_{\mathbf k}&=1+e^{-2i\mathbf{k}\cdot\mathbf{a}_1}+e^{i\mathbf{k}\cdot\mathbf{a}_2}+e^{-i\mathbf{k}\cdot(2\mathbf{a}_1+\mathbf{a}_2)}, \nn
    g_{\mathbf k}&=1+e^{-2i\mathbf{k}\cdot\mathbf{a}_1}-e^{i\mathbf{k}\cdot\mathbf{a}_2}-e^{-i\mathbf{k}\cdot(2\mathbf{a}_1+\mathbf{a}_2)}.
\label{eq:CS-params}
\end{align}

The magnon spectrum is obtained from the non-Hermitian matrix~\cite{xiao09}
\begin{align*}
    M_\mathbf{k}
    =
    \Sigma_3\, H_\mathbf{k},
    \quad
    \Sigma_3
    =
    \begin{pmatrix}
        \mathbb{I}_2 & 0 \\
        0 & -\mathbb{I}_2
    \end{pmatrix}.
\end{align*}
by solving the characteristic equation
\begin{align*}
    0
    =
    \det\!\left[
        E\,\mathbb{I}
        -
        M_\mathbf{k}
    \right]
    =
    (E - E_{1,\mathbf{k}})(E + E_{1,-\mathbf{k}})
    (E - E_{2,\mathbf{k}})(E + E_{2,-\mathbf{k}}) .
\end{align*}
Note that $E_{1,\mathbf{k}} - E_{1,-\mathbf{k}} + E_{2,\mathbf{k}} - E_{2,-\mathbf{k}}={\rm Tr}\,M_{\bf k}={\rm Tr}\,h_{\bf k}-{\rm Tr}\,h_{\bf k}^\top=0$, which implies $E_{\sigma,-{\bf k}}=E_{\sigma,{\bf k}}\,\forall\sigma$ as long as the bands do not cross each other. The corresponding eigenvectors form the columns of the matrix
$T_{\mathbf{k}}$, which diagonalizes $M_{\bf k}$ as
\begin{align}
    M_\mathbf{k}T_{\mathbf{k}} =
    T_{\mathbf{k}}
    \,
    \mathrm{diag}\!\left(
        E_{1,\mathbf{k}},\,
        E_{2,\mathbf{k}},\,
        -E_{1,-\mathbf{k}},\,
        -E_{2,-\mathbf{k}}
    \right).
\label{eq:Mdiag}
\end{align}
and represents the Bogoliubov transformation
$\Psi_{\mathbf{k}}
    =
    T_{\mathbf{k}}\,
    \Phi_{\mathbf{k}}$
to the quasiparticle basis
    $\Phi_{\mathbf{k}}
    =
    \begin{pmatrix}
        \alpha_{\mathbf{k}} & 
        \beta_{\mathbf{k}} &
        \alpha_{-\mathbf{k}}^\dagger &
        \beta_{-\mathbf{k}}^\dagger
    \end{pmatrix}^\top$.
Since the transformation must preserve the bosonic commutation relations, $[\Psi_{\mathbf{k}},\Psi_{\mathbf{k}}^\dagger]=[\Phi_{\mathbf{k}},\Phi_{\mathbf{k}}^\dagger]=\Sigma_3$,
$T_{\mathbf{k}}$ must satisfy the paraunitary condition $T_{\mathbf{k}}^\dagger \Sigma_3 T_{\mathbf{k}}=\Sigma_3$. This enables us to rewrite Eq. \eqref{eq:Mdiag} to
\begin{align*}
    T_{\mathbf{k}}^\dagger H_{\bf k} T_{\mathbf{k}}
    =\mathrm{diag}\!\left(
        E_{1,\mathbf{k}},\,
        E_{2,\mathbf{k}},\,
        E_{1,-\mathbf{k}},\,
        E_{2,-\mathbf{k}}
    \right),
\end{align*}
so that the magnon Hamiltonian is diagonalized as
\begin{align*}
    H_m
    &=
    \sum_{\mathbf{k}\in{\rm BZ}'}
    \left[
        \frac{1}{2}
        \Phi_{\mathbf{k}}^\dagger
        \mathrm{diag}\!\left(
        E_{1,\mathbf{k}},\,
        E_{2,\mathbf{k}},\,
        E_{1,-\mathbf{k}},\,
        E_{2,-\mathbf{k}}
        \right)
        \Phi_{\mathbf{k}}
        - \frac{1}{2}(\epsilon_A+\epsilon_B)
    \right]
    + E_0
    \nonumber\\
    &=
    \sum_{\mathbf{k}\in{\rm BZ}'}
    \left[E_{1,\mathbf{k}}\, \alpha_{\mathbf{k}}^\dagger \alpha_{\mathbf{k}}
    + E_{2,\mathbf{k}}\, \beta_{\mathbf{k}}^\dagger \beta_{\mathbf{k}}\right]
    + E_2.
\end{align*}
The $\mathbf{k}$-independent constant
\begin{align*}
    E_2
    =
    E_0
    +
    \sum_{\mathbf{k}\in{\rm BZ}'}
    \left[
        \frac{1}{2}\sum_{\sigma=1}^{2} E_{\sigma,\mathbf{k}}
        - \frac{1}{2}(\epsilon_A+\epsilon_B)
    \right]
\end{align*}
corrects the classical ground-state energy $E_0$ by the zero-point energy originating from the Bogoliubov
transformation together with the normal-ordering counterterm.
The excitation spectrum is thus fully characterized by the two positive-energy
magnon branches $E_{1,\mathbf{k}}$ and $E_{2,\mathbf{k}}$.

At finite temperature, the quasiparticle populations follow the Bose
distribution $n_{\sigma,{\bf k}}(T)=\lbrace\exp{[E_{\sigma,\mathbf{k}}/\kb T]}-1\rbrace^{-1}$. Using the Bogoliubov transformation \eqref{eq:CS-Bogol}, we obtain the thermal expectation values
\begin{align*}
    \langle a_{\mathbf{k}}^\dagger a_{\mathbf{k}} \rangle
    &=
    |(T_{\bf k})_{11}|^2\, \langle \alpha_{\bf k}^\dag \alpha_{\bf k} \rangle
    + |(T_{\bf k})_{12}|^2\, \langle \beta_{\bf k}^\dag \beta_{\bf k} \rangle
    + |(T_{\bf k})_{13}|^2\, \langle \alpha_{-\bf k} \alpha_{-\bf k}^\dag \rangle
    + |(T_{\bf k})_{14}|^2\, \langle \beta_{-\bf k} \beta_{-\bf k}^\dag \rangle \nn
    &=
    |(T_{\bf k})_{11}|^2\, n_{1,{\bf k}}
    + |(T_{\bf k})_{12}|^2\, n_{2,{\bf k}}
    + |(T_{\bf k})_{13}|^2\,[1+n_{1,-{\bf k}}]
    + |(T_{\bf k})_{14}|^2\,[1+n_{2,-{\bf k}}],\nn
    \langle b_{\mathbf{k}}^\dagger b_{\mathbf{k}} \rangle
    &=
    |(T_{\bf k})_{21}|^2\, \langle \alpha_{\bf k}^\dag \alpha_{\bf k} \rangle
    + |(T_{\bf k})_{22}|^2\, \langle \beta_{\bf k}^\dag \beta_{\bf k} \rangle
    + |(T_{\bf k})_{23}|^2\, \langle \alpha_{-\bf k} \alpha_{-\bf k}^\dag \rangle
    + |(T_{\bf k})_{24}|^2\, \langle \beta_{-\bf k} \beta_{-\bf k}^\dag \rangle \nn
    &=
    |(T_{\bf k})_{21}|^2\, n_{1,{\bf k}}
    + |(T_{\bf k})_{22}|^2\, n_{2,{\bf k}}
    + |(T_{\bf k})_{23}|^2\,[1+n_{1,-{\bf k}}]
    + |(T_{\bf k})_{24}|^2\,[1+n_{2,-{\bf k}}].
\end{align*}
The sublattice-resolved spin fluctuations at temperature $T=0$ are thus
\begin{align}
    \delta S_A(T=0)
    &=
    \frac{2}{N}
    \sum_{\mathbf{k}\in{\rm BZ}'}
    \langle a_{\mathbf{k}}^\dagger a_{\mathbf{k}} \rangle_{T=0}
    = \frac{2}{N}
    \sum_{\mathbf{k}\in{\rm BZ}'}
    \left[|(T_{\bf k})_{13}|^2 + |(T_{\bf k})_{14}|^2\right],\nn
    \delta S_B(T=0)
    &=
    \frac{2}{N}
    \sum_{\mathbf{k}\in{\rm BZ}'}
    \langle b_{\mathbf{k}}^\dagger b_{\mathbf{k}} \rangle_{T=0}
    =\frac{2}{N}
    \sum_{\mathbf{k}\in{\rm BZ}'}
    \left[|(T_{\bf k})_{23}|^2 + |(T_{\bf k})_{24}|^2\right].
\end{align}
For illustration, we quote the zero-temperature results obtained for
the parameter set $(h_z,J_\Gamma) = (0.2,1.0)J_z$,
\begin{align*}
    \delta S_A^{\rm HP} &\simeq 0.0128,
    \qquad
    \delta S_A^{\rm DMRG} \simeq 0.0162,
    \nn
    \delta S_B^{\rm HP} &\simeq 0.0187,
    \qquad
    \delta S_B^{\rm DMRG} \simeq 0.0197.
\end{align*}
all of which remain small compared to the classical value $S=1/2$, indicating that the system is well within the validity regime of HP theory.

To compute the QSSC of the CS phase, we start by rewriting Eq. \eqref{eq:chi_i} in terms of the sublattice index $i'$, 
\begin{align*}
    \chi = \frac{1}{2N} \sum_{i'} 
    \langle & \Sp^A_{i'} \cdot \Sp^A_{(i'_1,i'_2+1)} \times \Sp^B_{(i'_1-1,i'_2)}
    +\Sp^A_{i'} \cdot \Sp^A_{(i'_1,i'_2-1)} \times \Sp^B_{i'} \nn
    & +\Sp^B_{i'} \cdot \Sp^B_{(i'_1,i'_2+1)} \times \Sp^A_{i'}
    +\Sp^B_{i'} \cdot \Sp^B_{(i'_1,i'_2-1)} \times \Sp^A_{(i'_1+1,i'_2)} \rangle,
\end{align*}
where the first and second lines correspond to terms with $i\in A$ and $i\in B$ in Eq. \eqref{eq:chi_i}, respectively. Following the procedure of Eq. \eqref{eq:chi_vsc} and taking the average over each pair of edge-sharing triangles, we get
\begin{align}
    \chi = \frac{1}{4N}\sum_{i'} \langle \Sp^A-\Sp^B \rangle_{\rm cl} \cdot 
    \langle & \delta \Sp^A_{(i'_1,i'_2+1)} \times \delta \Sp^B_{(i'_1-1,i'_2)}
    + \delta \Sp^B_{(i'_1-1,i'_2)} \times \delta \Sp^A_{i'}
    + \delta \Sp^A_{(i'_1,i'_2-1)} \times \delta \Sp^B_{i'}
    + \delta \Sp^B_{i'} \times \delta \Sp^A_{i'} \nn
    & - \delta \Sp^B_{(i'_1,i'_2+1)} \times \delta \Sp^A_{i'}
    - \delta \Sp^A_{i'} \times \delta \Sp^B_{i'}
    - \delta \Sp^B_{(i'_1,i'_2-1)} \times \delta \Sp^A_{(i'_1+1,i'_2)}
    - \delta \Sp^A_{(i'_1+1,i'_2)} \times \delta \Sp^B_{i'}
    \rangle
\label{eq:chi_sublattice}
\end{align}
to the second order in $\delta \Sp$. Here,
\begin{align*}
    \langle \Sp^A-\Sp^B \rangle_{\rm cl} = \frac{1}{2} \begin{pmatrix}
        \frac{\sqrt{3}}{2}(\sin\Theta_A+\sin\Theta_B) \\
        \frac{1}{2}(\sin\Theta_A+\sin\Theta_B) \\
        \cos\Theta_A+\cos\Theta_B
    \end{pmatrix}
\end{align*}
is the difference between the classical spin configurations in sublattice $A$ and $B$. For the computation of $\chi$ within the free HP theory, the relevant component of the VSC is
\begin{align*}
    \langle \delta \Sp^A_{i'} \times \delta \Sp^B_{j'} \rangle = \begin{pmatrix}
        \big\langle \sin\Theta_A \tilde S_{i'}^{A,x} \big( \frac{1}{2}\cos\Theta_B \tilde S_{j'}^{B,x}-\frac{\sqrt{3}}{2}\tilde S_{j'}^{B,y} \big)
        -\sin\Theta_B \tilde S_{j'}^{B,x} \big( \frac{1}{2}\cos\Theta_A \tilde S_{i'}^{A,x}+\frac{\sqrt{3}}{2}\tilde S_{i'}^{A,y} \big) \big\rangle \\ 
        \big\langle \big( \frac{\sqrt{3}}{2}\cos\Theta_A \tilde S_{i'}^{A,x}-\frac{1}{2}\tilde S_{i'}^{A,y} \big) \sin\Theta_B \tilde S_{j'}^{B,x} 
        - \big( \frac{\sqrt{3}}{2}\cos\Theta_B \tilde S_{j'}^{B,x}+\frac{1}{2}\tilde S_{j'}^{B,y} \big) \sin\Theta_A \tilde S_{i'}^{A,x} \big\rangle \\ 
        \big\langle \big( \frac{\sqrt{3}}{2}\cos\Theta_A \tilde S_{i'}^{A,x}-\frac{1}{2}\tilde S_{i'}^{A,y} \big) \big( \frac{1}{2}\cos\Theta_B \tilde S_{j'}^{B,x}-\frac{\sqrt{3}}{2}\tilde S_{j'}^{B,y} \big) 
        - \big( \frac{\sqrt{3}}{2}\cos\Theta_B \tilde S_{j'}^{B,x}+\frac{1}{2}\tilde S_{j'}^{B,y} \big) \big( \frac{1}{2}\cos\Theta_A \tilde S_{i'}^{A,x}+\frac{\sqrt{3}}{2}\tilde S_{i'}^{A,y} \big) \big\rangle
    \end{pmatrix},
\end{align*}
which yields
\begin{align}
    \langle \Sp^A-\Sp^B \rangle_{\rm cl} \cdot \langle \delta \Sp^A_{i'} \times \delta \Sp^B_{j'} \rangle =& -\frac{1}{2}(\sin\Theta_A+\sin\Theta_B)\langle \sin\Theta_A \tilde S_{i'}^{A,x} \tilde S_{j'}^{B,y} + \sin\Theta_B \tilde S_{i'}^{A,y} \tilde S_{j'}^{B,x} \rangle \nn
    & -\frac{1}{2}(\cos\Theta_A+\cos\Theta_B)\langle \cos\Theta_A \tilde S_{i'}^{A,x} \tilde S_{j'}^{B,y} + \cos\Theta_B \tilde S_{i'}^{A,y} \tilde S_{j'}^{B,x} \rangle \nn
    =& -\frac{1}{2}[1+\cos(\Theta_A-\Theta_B)]\langle \tilde S_{i'}^{A,x} \tilde S_{j'}^{B,y} + \tilde S_{i'}^{A,y} \tilde S_{j'}^{B,x} \rangle \nn
    =& \frac{i}{4} [1+\cos(\Theta_A-\Theta_B)] \langle a_{i'}b_{j'}-a^\dag_{i'}b^\dag_{j'} \rangle \nn
    =& \frac{i}{2N} [1+\cos(\Theta_A-\Theta_B)] \sum_{\mathbf{k}\in{\rm BZ}'} e^{i\mathbf{k}\cdot \mathbf{\delta}}\langle a_\mathbf{k} b_{-\mathbf{k}} - a^\dag_\mathbf{k} b^\dag_{-\mathbf{k}} \rangle,
\label{eq:chi_CS_summand}    
\end{align}
where $\mathbf{\delta}=\mathbf{R}_{i'}-\mathbf{R}_{j'}$. In the last line, we used that, due to the form of the Hamiltonian \eqref{eq:H_CS}, $\langle a_\mathbf{k} b_\mathbf{q} \rangle$ is nonzero only if $\mathbf{k}=-\mathbf{q}$. Finally, plugging Eq. \eqref{eq:chi_CS_summand} into Eq. \eqref{eq:chi_sublattice} gives
\begin{align}
    \chi =& \frac{i}{8N} [1+\cos(\Theta_A-\Theta_B)] \sum_{\mathbf{k}\in{\rm BZ}'} \left[ (-1-e^{2i\mathbf{k}\cdot\mathbf{a}_1}+e^{-i\mathbf{k}\cdot\mathbf{a}_2}+e^{i\mathbf{k}\cdot(2\mathbf{a}_1+\mathbf{a}_2)})\langle a_\mathbf{k} b_{-\mathbf{k}} \rangle - {\rm c.c.} \right] \nn
    =& \frac{-i}{8N} [1+\cos(\Theta_A-\Theta_B)] \nn
    & \times \sum_{\mathbf{k}\in{\rm BZ}'} \Big[g_{-\bf k} \big\{(T_{\bf k})_{11}(T_{\bf k})_{41}^*\,(1+n_{1,{\bf k}})+(T_{\bf k})_{12}(T_{\bf k})_{42}^*\,(1+n_{2,{\bf k}})+(T_{\bf k})_{13}(T_{\bf k})_{43}^*\,n_{1,{-\bf k}}+(T_{\bf k})_{14}(T_{\bf k})_{44}^*\,n_{2,{-\bf k}}\big\} - {\rm c.c.}\Big].
\end{align}
In the last line, $g_{-\bf k}$ is as defined in Eq. \eqref{eq:CS-params} and the Bogoliubov transformation \eqref{eq:CS-Bogol} was applied to evaluate $\langle a_\mathbf{k} b_{-\mathbf{k}} \rangle$.
\end{widetext}

\end{document}